\documentclass[apj, revtex4]{emulateapj}

\usepackage{epsf}
\usepackage{color}
\usepackage{amsmath}
\usepackage{graphicx}
\usepackage[colorlinks,urlcolor=blue,citecolor=blue,linkcolor=blue]{hyperref}
\usepackage{threeparttable}
\usepackage{multirow}
\usepackage{natbib}
\usepackage{etoolbox}

\pdfoutput=1

\newcommand{\hst}{\textit{HST}}

\newcommand{\acsb}{\hbox{$\mathrm{B}_{435}$}}
\newcommand{\acsv}{\hbox{$\mathrm{V}_{606}$}}
\newcommand{\acsi}{\hbox{$i_{775}$}}
\newcommand{\acsz}{\hbox{$z_{850}$}}

\newcommand{\wfcy}{\hbox{$\mathrm{Y}_{105}$}}
\newcommand{\wfcj}{\hbox{$\mathrm{J}_{125}$}}
\newcommand{\wfch}{\hbox{$\mathrm{H}_{160}$}}

\newcommand{\Msol}{\hbox{$\mathrm{M}_\odot$}}
\newcommand{\msol}{\hbox{$\mathrm{M}_\odot$}}
\newcommand{\AAA}{\hbox{\AA}~}

\newcommand{\kms}{\hbox{km~s$^{-1}$}}

\newcommand{\mpc}{\hbox{Mpc$^{-1}$}}

\newcommand{\um}{\hbox{$\mu$m}}

\newcommand{\etal}{et al.}
\newcommand{\eg}{e.g.}
\newcommand{\ie}{i.e.}

\makeatletter
\patchcmd{\NAT@citex}
  {\@citea\NAT@hyper@{%
     \NAT@nmfmt{\NAT@nm}%
     \hyper@natlinkbreak{\NAT@aysep\NAT@spacechar}{\@citeb\@extra@b@citeb}%
     \NAT@date}}
  {\@citea\NAT@nmfmt{\NAT@nm}%
   \NAT@aysep\NAT@spacechar\NAT@hyper@{\NAT@date}}{}{}

\patchcmd{\NAT@citex}
  {\@citea\NAT@hyper@{%
     \NAT@nmfmt{\NAT@nm}%
     \hyper@natlinkbreak{\NAT@spacechar\NAT@@open\if*#1*\else#1\NAT@spacechar\fi}%
       {\@citeb\@extra@b@citeb}%
     \NAT@date}}
  {\@citea\NAT@nmfmt{\NAT@nm}%
   \NAT@spacechar\NAT@@open\if*#1*\else#1\NAT@spacechar\fi\NAT@hyper@{\NAT@date}}

\shorttitle{}
\shortauthors{BOADA ET AL.}

\slugcomment{Accepted for Publication in the Astrophysical Journal}

\begin{document}

\title{The Role of Bulge Formation in the Homogenization of Stellar Populations at $z\sim2$ as revealed by Internal Color Dispersion in CANDELS}

\author{\sc Steven Boada\altaffilmark{1}, 
V.~Tilvi\altaffilmark{1},
C.~Papovich\altaffilmark{1},
R.F.~Quadri\altaffilmark{1}
M.~Hilton\altaffilmark{2,}\altaffilmark{3},
S.~Finkelstein\altaffilmark{4},
Yicheng Guo\altaffilmark{8},
N.~Bond\altaffilmark{5}
C.~Conselice\altaffilmark{3},
A.~Dekel\altaffilmark{10}
H.~Ferguson\altaffilmark{6},
M.~Giavalisco\altaffilmark{7}
N.A.~Grogin\altaffilmark{6},
D.D.~Kocevski\altaffilmark{9},
A.M. Koekemoer\altaffilmark{6},
D.C.~Koo\altaffilmark{8}
} 

\altaffiltext{1}{George P.\ and Cynthia Woods Mitchell Institute for
Fundamental Physics and Astronomy, and Department of Physics and Astronomy,
Texas A\&M University, College Station, TX, 77843-4242;
boada@physics.tamu.edu}
\altaffiltext{2}{Centre for Astronomy \& Particle Theory, School of Physics \& Astronomy, 
University of Nottingham, Nottingham, NG7 2RD, UK}
\altaffiltext{3}{Astrophysics \& Cosmology Research Unit, School of Mathematics, Statistics \& 
Computer Science, University of KwaZulu-Natal, Durban 4041, South Africa}
\altaffiltext{4}{Department of Astronomy, The University of Texas at Austin, Austin, TX 78712}
\altaffiltext{5}{Laboratory for Observational Cosmology, Astrophysics Science Division, Code 665, Goddard Space Flight Center, Greenbelt, MD 20771}
\altaffiltext{6}{Space Telescope Science Institute, 3700 San Martin Drive, Baltimore, MD 21218}
\altaffiltext{7}{Department of Astronomy, University of Massachusetts, 710 North Pleasant Street, Amherst, MA 01003}
\altaffiltext{8}{UCO/Lick Observatory \& Department of Astronomy and Astrophysics, University of California, Santa Cruz, CA 95064}
\altaffiltext{9}{Department of Physics and Astronomy, University of Kentucky, Lexington, KY 40506}
\altaffiltext{10}{Center for Astrophysics and Planetary Science, Racah Institute of Physics, The Hebrew University, Jerusalem 91904, Israel}

\begin{abstract} \noindent We use data from the Cosmic Assembly Near-infrared Deep
Extragalactic Legacy Survey to study how the spatial variation in the stellar populations of
galaxies relate to the formation of galaxies at $1.5 < z < 3.5$. We use the Internal Color
Dispersion (ICD), measured between the rest-frame UV and optical bands, which is sensitive
to age (and dust attenuation) variations in stellar populations. The ICD shows a relation
with the stellar masses and morphologies of the galaxies. Galaxies with the largest
variation in their stellar populations as evidenced by high ICD have disk-dominated
morphologies (with S\'{e}rsic indexes $< 2$) and stellar masses between $10 <
\mathrm{Log~M}/ \Msol< 11$. There is a marked decrease in the ICD as the stellar mass and/or
the S\'ersic index increases. By studying the relations between the ICD and other galaxy
properties including sizes, total colors, star-formation rate, and dust attenuation, we
conclude that the largest variations in stellar populations occur in galaxies where the
light from newly, high star-forming clumps contrasts older stellar disk populations. This
phase reaches a peak for galaxies only with a specific stellar mass range, $10 <
\mathrm{Log~M}/ \Msol < 11$, and prior to the formation of a substantial bulge/spheroid. In
contrast, galaxies at higher or lower stellar masses, and/or higher S\'{e}rsic index ($n >
2$) show reduced ICD values, implying a greater homogeneity of their stellar populations.
This indicates that if a galaxy is to have both a quiescent bulge along with a star forming
disk, typical of Hubble Sequence galaxies, this is most common for stellar masses $10 <
\mathrm{Log~M}/\Msol < 11$ and when the bulge component remains relatively small ($n<2$).
\end{abstract}


\section{INTRODUCTION}\label{sec:introduction} In the local universe, galaxies separate into
two broad classes \citep{Kauffmann2003b, Baldry2004}: (1) large disk-dominated galaxies show
ongoing star formation rates (SFRs) comparable to their past averages as evidenced by their
overall blue colors, and frequent spiral arms, and (2) spheroidal, or bulge-dominated
galaxies with current SFRs much less than their past averages. However, especially in the
more distant universe, there exists a third class of galaxies which exhibits a large amount
of stellar population diversity and irregular morphologies (\eg, \citealt{Giavalisco1996,
Lowenthal1997, Dickinson2000, Papovich2005}). The mechanism by which a galaxy transitions
from a uniformly blue, star forming galaxy (with either disk-dominated or irregular
morphologies) into a bulge-dominated, quiescent galaxy remains an open question.

Many studies have added to a growing body of evidence that the irregular morphologies are
due, at least in part, to heterogeneous (clumpy) star formation (\eg, \citealt{Adamo2013,
Elmegreen2005, Elmegreen2008, Elmegreen2009, Elmegreen2009a, Guo2012, Wuyts2011, Wuyts2012,
Wuyts2013, Guo2014}). Believed to be formed through gravitational instabilities in gas-rich
disks (\eg, \citealt{Keres2005, Dekel2006, Dekel2009a, Dekel2014}), these clumps typically
contribute $10-20\%$ of the total galaxy light \citep{ForsterSchreiber2011} and most have
bluer colors and elevated specific star formation rates (sSFR) compared to the surrounding
regions \citep{Guo2012, Guo2014}. This contrast between a young star forming region and the
older underlying galaxy population may lead to a large amount of variation in the internal
colors of a galaxy.

The ability to investigate the separation (heterogeneity) of the different stellar
populations (blue star forming clumps versus the older, redder stellar disk) becomes
increasingly difficult at high redshift, because a galaxy's morphology often depends heavily
on the bandpass in which the galaxy is observed. For example, observed optical light
(rest-frame UV) traces the distribution of unobscured star-forming regions
\citep{Dickinson2000} which are often non-uniformly distributed.

Recent observations of clumpy galaxies (\eg, \citealt{Guo2012, Wuyts2012, Wuyts2013,
Adamo2013, Guo2014}) find that clumps located toward galaxy centers are older and redder
than clumps located toward the outskirts. This suggests that highly star forming clumps may
migrate toward the center where they could condense to form a psuedo-bulge (\eg,
\citealt{Ceverino2010}). The formation of such a bulge, in turn, could be sufficient to
stabilize the stellar disk against new star-formation \citep{Martig2009}, homogenizing the
stellar populations and reducing the variations in internal color. If this scenario is
correct, we would expect to observe a relation between the internal color dispersion of a
galaxy and the formation of a bulge.

To investigate any potential relation, we require a tool to quantify the stellar population
diversity of a galaxy. Traditional morphological classification methods: visual inspection,
S\'{e}rsic profile fitting, concentration and asymmetry values \citep{Conselice2004}, and
Gini coefficients \citep{Lotz2004a} are calculated in a single bandpass and, as such, do not
provide enough insight into the stellar population variations. \cite{Papovich2003} developed
a differential morphological measure, the Internal Color Dispersion (ICD), which quantifies
the heterogeneity of stellar populations (such as with clumps) between bandpasses; galaxies
with non-uniform distributions of young stellar populations give rise to a larger ICD than a
galaxy with a homogeneous or evenly mixed stellar population. Similarly, galaxies with
variable amounts of extinction will also exhibit large ICD values, due to the spatial
variations in the reddening. Combining the ICD with other morphological indicators that
quantify the distribution of light in a galaxy (\eg\ S\'{e}rsic, Gini-$M_{20}$,
concentration and asymmetry) provides insight into galaxies physical properties such as
clumpy star-formation and their evolution in both overall color and structure.

In this paper, we build upon previous studies (\eg, \citealt{Papovich2003, Papovich2005,
Bond2011, Law2012}) and we use the ICD as the primary metric in a comparative study of the
rest-frame UV and optical morphological properties of the largest sample yet ($>800$) of
$1.5 < z < 3.5$ galaxies to date. In addition to broader IR coverage, we combine optical and
near-IR data from \hst\ from the Southern region of The Great Observatories Origins Deep
Survey and Cosmic Assembly Near-Infrared Deep Extragalactic Survey, which also includes the
deeper data in the Hubble Ultra Deep Field. We investigate the presence of heterogeneous
stellar populations in a sample of $1.5 < z < 3.5$ galaxies and interpret our results in the
context of galaxy evolution where a galaxy transitions from a disk-dominated to a more
bulge-dominated morphology while experiencing a suppression of the diversity in its stellar
populations.

In Section~\ref{sec:data} we discuss the data and sample selection. We discuss the ICD
measuring technique in detail in Section~\ref{sec:the ICD}. We compare ICD and other galaxy
properties in Section~\ref{sec:results}, and discuss possible causes of high ICD in
Section~\ref{sec:high icd}. We discuss the role of star-forming clumps in galaxy evolution
in Section~\ref{sec:On Morphology}, and finally summarize our results in
Section~\ref{sec:summary}.

Throughout this paper, we use a concordance cosmological model ($\Omega_\Lambda = 0.7$,
$\Omega_m = 0.3$, and $H_0= 70$ \kms \mpc), assume a Chabrier initial mass function (IMF;
\citealt{Chabrier2003}), and use AB magnitudes \citep{Oke1974}.

\section{Data}\label{sec:data} For this work, we make use of deep imaging data taken as part
of the Cosmic Assembly Near-Infrared Deep Extragalactic Survey (CANDELS; PIs Faber and
Ferguson; \citealt{Koekemoer2011, Grogin2011}) of the Southern region of The Great
Observatories Origins Deep Survey (GOODS, hereafter referred to as GOODS-S,
\citealt{Giavalisco2004}) along with extremely deep observations of the Hubble Ultra Deep
Field (HUDF). The large variation in depth between the GOODS-S observations and the HUDF
ensures that we have access to galaxies with a large range in stellar mass.

We take advantage of the third version of the Advanced Camera for Surveys (ACS) mosaics in
GOODS-S which includes images taken with F435W, F606W, F775W, and F850LP (hereafter referred
to as \acsb, \acsv, \acsi, and \acsz). In addition, we use imaging obtained from Wide Field
Camera 3 (WFC3) in F098M, F105W, F125W, and F160W (hereafter as Y$_{098}$, \wfcy, \wfcj, and
\wfch). The WFC3 imaging is split into two parts; the deep portion covers the central 50\%
of GOODS-S with the wide portion covering the Southern-most 25\% to a depth approximately 1
magnitude shallower than the deep portion. To complete our WFC3 coverage of GOODS-S, we make
use of the WFC3 Science Organizing Committee's Early Release Science program (PI O'Connell;
\citealt{Windhorst2011}) which covers the Northern 25\% in Y$_{098}$, \wfcj, and \wfch.

In addition, we include very deep imaging of the HUDF taken with both ACS
\citep{Beckwith2006} and WFC3 as part of HUDF09 (PI Illingsworth; \eg, \citealt{Bouwens2010,
Oesch2010}) and UDF12 (PI Ellis; \citealt{Ellis2013}). Both the ACS and WFC3 datasets
contain imaging using the previously mentioned seven bandpasses.

\subsection{Source Detection and Photometric Data} We use a multi-wavelength photometric
catalog of the GOODS-S region provided by \cite{Guo2013}. In this catalog, the sources are
detected in a \wfch\ mosaic drizzled to $0.06''/$pixel. Along with the source detections,
photometry covering 0.4--8~\micron\ (VIMOS U, \acsb, \acsv, \acsi, \acsz, Y$_{098}$, \wfcy,
\wfcj, \wfch, ISSAC Ks, IRAC 3.6\micron, 4.5\micron, 5.8\micron, and 8\micron) is provided.
The photometric data for the \textit{HST} bands are computed by {\tt SExtractor}
\citep{Bertin1996} on images which have had their point spread function (PSF) convolved to
match the \wfch\ resolution, while the lower resolution ground based and
\textit{Spitzer}/IRAC imaging is processed by {\tt TFIT} \citep{Laidler2006}. See
\cite{Guo2013} for full details on the construction of the photometric catalog.

\subsection{Photometric Redshifts}\label{sec:photoz} We make use of the photometric
redshifts for the galaxies in our sample derived by \cite{Dahlen2013}. This photometric
redshift catalog is constructed by combining redshifts derived by 11 different investigators
using different photometric redshift codes e.g., {\tt EAZY} \citep{Brammer2008}, {\tt
HyperZ} \citep{Bolzonella2000}, {\tt LePhare} \citep{Arnouts1999, Ilbert2006}, etc. These
photometric redshift codes derive redshifts by fitting different combination of spectral
energy distribution templates and priors including available spectroscopic redshifts. We
make use of spectroscopic redshifts where available (see \citealt{Dahlen2013} for the
compilation of spectroscopic datasets). The scatter between the photometric redshifts and
corresponding spectroscopic redshifts, which span $z\sim0-6$, is rms$[\Delta z/(1+z_{\rm
spec})] = 0.03$ where $\Delta z= z_{\rm spec} - z_{\rm phot}$.

\subsection{Sample Selection}\label{sec:QC and selection} \cite{Papovich2003} showed that
galaxies with heterogeneous stellar populations have high ICD between their rest-frame UV
and optical light. For our galaxy sample with $1.5 < z < 3.5$, the CANDELS ACS and WFC3
bands probe rest-frame UV and optical wavelengths respectively. Using the photometric
redshifts described previously (Section \ref{sec:photoz}) we select all galaxies with
photometric or spectroscopic redshifts in this range. We apply a magnitude limit of 25 mag
in \wfch, which allows for a robust determination of morphological parameters and the ICD
values (see Section \ref{sec:ICD vs ston} for reasoning). To avoid stars, we remove objects
with stellarity index $> 0.78$; stellarity derived using {\tt SExtractor} gives a likelihood
of an object being a star. Applying this set of criteria yields a parent sample 3369
galaxies.

\subsection{Stellar Masses}\label{sec:Stellar Masses} Because the photometric redshifts are
the result of the combination of of the probability density functions from many routines,
\cite{Dahlen2013} show they provide more accurate estimates of the redshift. The method
employed by \cite{Dahlen2013} does not provide additional physical properties for the
CANDELS galaxies. We derive stellar population parameters for the galaxies in our sample
using {\tt FAST} \citep{Kriek2009}, a stellar population synthesis modeling and fitting
code. We fix the galaxy's redshift at the photometric redshift from the \cite{Dahlen2013}
catalog and fit 14-band galaxy photometry covering $0.4-8$ \micron\ with model spectral
energy distributions to estimate the stellar masses (along with dust attenuation) for the
galaxies in the sample. We used models for a range of stellar population properties from the
2003 version of the Bruzual \& Charlot models (\citealt{Bruzual2003}) stellar population
synthesis models allowing for a range of attenuation (${\rm A_{v}} = 0-4$ mag) using the
\cite{Calzetti2000} extinction law. We opt to use \cite{Bruzual2003} to facilitate the
comparison to other studies and because recent work shows that the Bruzual \& Charlot 2007
models may overestimate the contributions of late type giants to the near-IR (\eg,
\citealt{Zibetti2012}). We assumed models with a delayed, exponentially declining (Log $\tau
= 7-11$) star formation history, solar metallicity and a \cite{Chabrier2003} IMF.

Using a Salpeter IMF \citep{Salpeter1955} would to first order increase systematically the
stellar masses by $\simeq0.25$ dex. Using different assumptions for the stellar population
metallicities will affect the derived stellar masses by $\simeq0.2$ dex
\citep{Papovich2001}. The average uncertainty in stellar mass for our sample is 0.11 dex,
although errors from systematics in the metallicities, star-formation histories and other
aspects of the spectral energy distribution fitting likely dominate (\eg,
\citealt{Kriek2009, Papovich2014a}).

The errors on the photometric redshifts do not influence greatly our conclusions. While,
photometric error uncertainties can affect the derivation of stellar population parameters,
for small photometric redshift errors, like those for our sample, the contribution to the
stellar mass uncertainty is minimal because small changes in redshift translate to small
changes in distance (see discussion in \citealt{Dickinson2003a}). For example, our
uncertainty of rms$[\Delta z/(1+z)] = 0.03$ corresponds to a change in luminosity distance
of $\sim$5\% for our sample. Because the stellar mass scales quadratically with luminosity
distance, the mass uncertainty from photometric redshift errors is similarly small,
$\sim$10\%. This error is much smaller than systematic uncertainties in stellar mass arising
from model fitting (see discussion in \cite{Papovich2014a}), and so we neglect errors from
the photometric redshifts on the stellar masses.

Along with stellar masses {\tt FAST} provides an estimate of the total attenuation ($\rm
A_V$) due to dust in a galaxy. We use this value in Section~\ref{sec:attenuation} to
understand better how the ICD might be effected by reddening due to dust.

\subsection{Morphological Indicators}\label{sec:morphdata} We use the structural catalog
provided by \cite{VanderWel2012} which provides S\'{e}rsic index, $n$, \citep{Sersic1963}
and the effective radius, $R_{e}$ (measured along the major axis of the galaxy), for all
sources in the GOODS-S catalog from \cite{Guo2013}. These parameters are produced by {\tt
GALFIT} \citep{Peng2002} using the best-fitting S\'{e}rsic models for the objects in the
\wfch\ imaging.

We also use the non-parametric indicators Gini parameter ($G$) and the second-order moment
of the brightest 20\% of the galaxy light, $M_{20}$. These indicators quantify the spatial
distribution of light without assuming a specific functional form when classifying the
morphologies of disturbed or irregular galaxies. For a complete description of this method
see \cite{Lotz2004a}.

\subsection{Color Gradients}\label{sec:data_colorgraident} We use color gradient
measurements from Hilton \etal\ (in prep), where Hilton et al. compute the color gradients
between the core and outskirts by centering annular apertures on each galaxy. They define
concentric, elliptical, annular apertures such that the total \wfch\ light of each galaxy is
divided equally between three annuli. The galaxy core corresponds to the ``inner'' aperture
and the galaxy outskirts correspond to the ``outer'' aperture. The color gradient of the
galaxy is denoted by \begin{equation} \Delta(m_1 - m_2) = (m_1 - m_2)_{\rm outer} - (m_1 -
m_2)_{\rm inner} \end{equation} where, $m_1$ and $m_2$ are the respective apparent
magnitudes. A galaxy where the core has a redder color than the outskirts will have a
negative color gradient and a galaxy with a bluer core than outskirts will have a positive
gradient.

\subsection{Star Formation Rates}\label{sec:data_sfr} In this work, we measure the total
star-formation rate (SFR) as the sum of the SFR implied by the observed UV luminosity
(uncorrected for dust extinction, $\rm SFR_{UV}$) and by the IR (as estimated from the
24\um\ luminosity, $\rm SFR_{IR}$). We convert the observed 24\um\ luminosity into a SFR
using the empirical relationship from \cite{Wuyts2011a}. For galaxies which are not detected
in 24\um, the SFRs are based on an attenuation-corrected UV flux using attenuation value
estimated by {\tt FAST}, so $\rm SFR_{Total}= SFR_{UV} \times \mathrm{Correction}$. In both
instances, we follow \cite{Kennicutt1998} to convert the luminosity to SFR. The
\cite{Kennicutt1998} conversion is Salpeter-based, whereas the masses and other derived
quantities are Chabrier-based. To account for the difference in IMF, we divide the total SFR
by 1.66, $\rm SFR_{Total}=(SFR_{UV} + SFR_{IR})/1.66$

\subsection{Discrete Star Forming Regions}\label{sec:data_clumps} In
Section~\ref{sec:clumps} we compare the internal color dispersion measurements to the
frequency of discrete star forming regions (colloquially ``clumps'') in galaxies using the
clump catalog of \cite{Guo2014}. In order to facilitate clump finding, \cite{Guo2012}
require galaxies to be approximately face-on (b/a $> 0.5$), reasonably extended ($R_e >
0.2''$), and with a sufficient brightness ($\wfch < 24.5$ mag).

Individual clumps are selected from the UV luminosity contrast between the individual clump
and the total galaxy. From the catalog, we select clumps with UV luminosity $>1\%$ of the
total galaxy UV light ($L_{clump}/L_{galaxy} > 0.01$) contained within galaxies in our
sample. While \cite{Guo2014} selects clumps with UV luminosity $>8\%$; we lower this
threshold to include more UV-faint clumps as they are redder and provide a more
comprehensive view of the ICD. This provides clump measurements for 180 objects, containing
between zero and eight clumps per galaxy.

\section{THE INTERNAL COLOR DISPERSION}\label{sec:the ICD} \subsection{Definition} First
developed by \cite{Papovich2003}, the Internal Color Dispersion (ICD) is a flux-independent
statistic, defined as the ratio of the squared difference of the image flux-intensity values
about the mean galaxy color to the square of the total image flux, or, \begin{align}
\xi(I_1,I_2) \equiv & \frac{ \sum\limits_{i=1}^{N_{pix}} ( I_{2,i} - \alpha\, I_{1,i} -
\beta)^2 - \sum\limits_{i=1}^{N_{pix}}(B_{2,i} - \alpha\, B_{1,i})^2 }
{\sum\limits_{i=1}^{N_{pix}} (I_{2,i}-\beta)^2 - \sum\limits_{i=1}^{N_{pix}}
(B_{2,i}-\alpha\, B_{1,i})^2} \nonumber\\ &\times 100\%. \label{eqn:xi} \end{align} The ICD
assesses the morphological difference of a galaxy between two passbands. The rest-frame
UV-optical ICD quantifies the spatial homogeneity (or lack thereof) between young stellar
populations (dominating the rest-frame UV light) and older stellar populations, contributing
most of the rest-frame optical light. Galaxies with spatially segregated stellar populations
with different colors would have large ICD (see \citealt{Papovich2003}). Whereas, a galaxy
with a homogeneous stellar population throughout would have a low ICD.

The terms in Equation~\ref{eqn:xi} sum over $i=1...N_{pix}$ with $N_{pix}$ being the number
of pixels associated with each galaxy, $I_1$ and $I_2$ are the image flux--intensity values
for each object obtained in each of two passbands and $B_1$ and $B_2$ are the pixel
intensity values from a contiguous, similarly sized, region of the two passbands not
associated with any object.

To determine which pixels belong to a galaxy we define the aperture size as $r_p(\eta =
0.2)$, where $r_p$ is the Petrosian radius of $I_2$ as defined for the \textit{Sloan Digital
Sky Survey} \citep{Blanton2001a}, and $\eta(r)$ is defined as the ratio of the galaxy
surface brightness, $I(r)$, averaged over an annulus of radius $r$, to the mean surface
brightness within this radius, $\langle I(r) \rangle$. The Petrosian radius depends on the
surface--brightness distribution of the galaxy and is thus independent of redshift or
systematics in the image calibration. The chosen aperture encompasses most of the light from
the galaxy without adding excessive background light. Therefore, the summations are over the
individual pixels associated with the object or blank region of a similar size.

The scaling factor $\alpha$ is the ratio of total fluxes between images $I_2$ and $I_1$,
while the linear offset $\beta$ adjusts (if necessary) for differences in the relative
background levels of the two images. In practice, we calculate these terms by minimizing the
statistic, $\chi^2 = \sum [ (I_2 - \alpha\, I_1 - \beta) / \sigma ]^2$, adjusting $\alpha$
and $\beta$ as free parameters (where $\sigma$ represents the uncertainties on the $I_2$
flux--intensity values).

The treatment of the background terms ($B_1$ and $B_2$) are important to the ICD, as the
background itself can contain an intrinsic amount of internal color dispersion based on our
definition. Therefore, to eliminate ICD contribution from the background, $B_1$ and $B_2$
are selected for each individual galaxies to contain the same number of pixels ($N_{pix}$)
at random locations excluding any preexisting objects, and include the same pixels in both
image 1 and 2. In order to minimize the ICD variations due to background, we select the
median ICD value computed from using nine different background regions. However, simple
fluctuations may not be the only source of ICD. Because, the WFC3 imaging PSF has been
convolved (see Section~\ref{sec:data}) to match that of the ACS, there exists the
possibility of correlated noise in the background measurements. This could have an effect if
the background region is not selected in the same way as the galaxy (\eg, choosing
individual pixels at random versus a single contiguous region). We test for possible effects
by choosing individual pixels not associated with an object at random from the imaging,
choosing a single contiguous \textit{square} region and contiguous \textit{circular} region.
We find no effect based on the morphology of chosen background pixels.

\subsection{ICD Dependence on Signal-to-Noise}\label{sec:ICD vs ston} Previous studies
\citep{Papovich2003, Papovich2005, Bond2011} noted that the ICD produces unreliable results
for galaxies with low signal-to-noise (S/N). We test the behavior of the ICD using simulated
galaxies with \textit{a priori} known (fixed) total flux and ICD. Each simulated galaxy is
placed throughout the \textit{HST} images at a random location and is then checked to insure
no overlap with an actual object by comparing the location with the {\tt SExtractor}
segmentation image. The flux of the galaxy is slowly decreased to a level barely above the
background. We then derive an ICD value for each simulated object using the definition in
Equation~\ref{eqn:xi}.

\begin{figure} \includegraphics[width=0.49\textwidth]{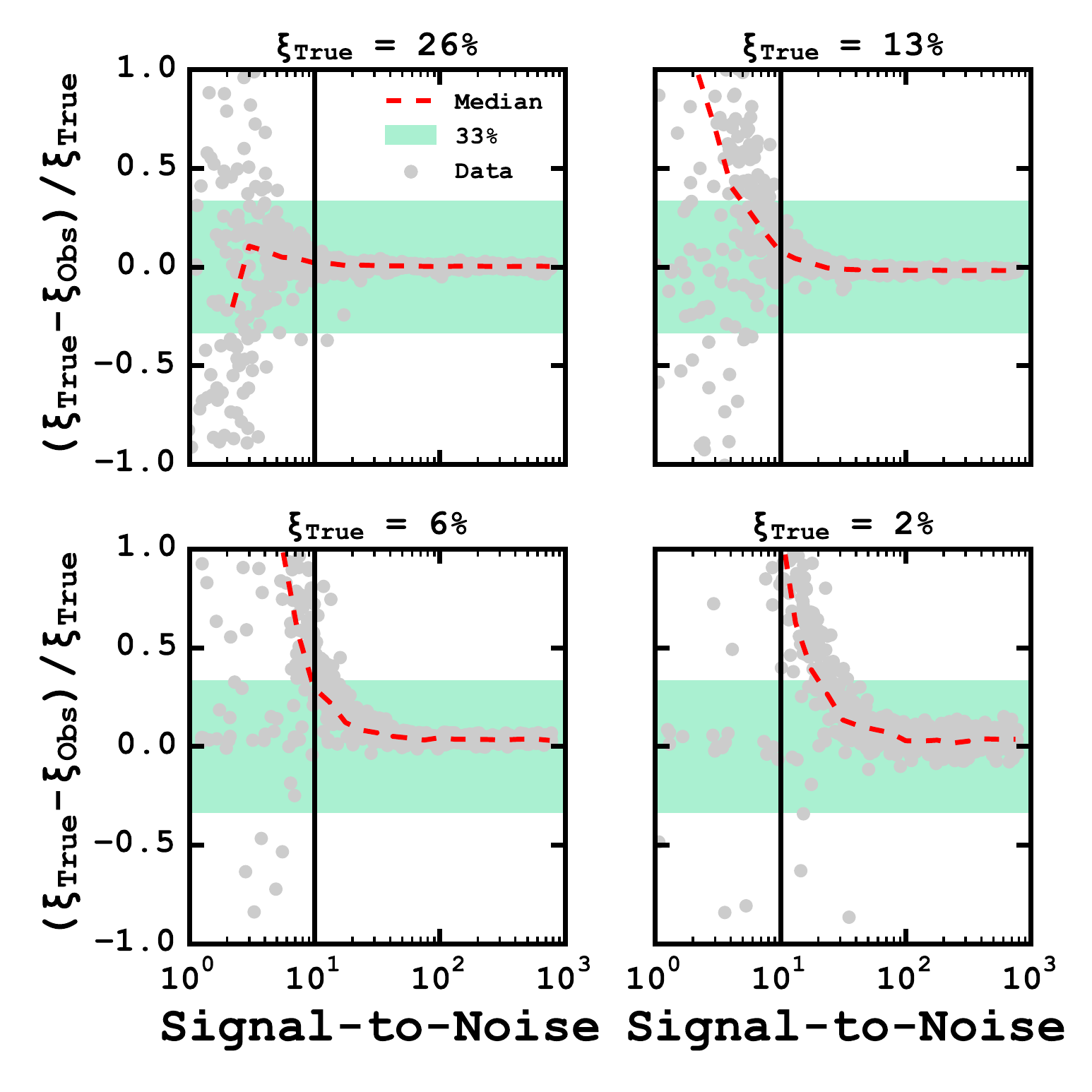}\caption{The fractional change of the Internal Color Dispersion as a function of signal-to-noise for two different simulated
objects. Gray points show individual S/N measurements. Red dashed lines show the median
value. Green shaded regions shows the area where the fractional change is less than 33\%
from the true value and serves as an aid to gauge accuracy. For the low ICD galaxy (top
panel) we find that galaxies with $\mathrm{S/N} < 10$ in the \textit{bluer band} the median
ICD deviates more than 34\% away from the true value. Therefore, we remove any object from
our initial sample that have $\mathrm{S/N} \geq 10$ in the bluer band. The S/N requirement
for galaxies with intrinsically high ICD could be set to a lower threshold, but we choose to
constant criterion to keep the selection uniform.} \label{fig:ston_I} \end{figure}

Figure~\ref{fig:ston_I} shows the fractional change (true minus observed divided by true) in
the ICD as a function of $\mathrm{S/N} = \sum f / \sqrt{\sum \sigma^2}$, where $f$
represents the flux--intensity values, and $\sigma^2$ is the image variance due to the noise
of the background. The sum is taken over all pixels identified as belonging to the object,
by computing the Petrosian radius and comparing the enclosed pixels to the {\tt SExtractor}
segmentation image. The four simulated objects illustrate how different levels of ICD (2\%,
6\%, 13\%, and 26\%) are effected by the galaxy's S/N.

The behavior of the fractional change is dramatically different depending on whether a
galaxy has intrinsically ``low'' or ``high'' ICD. For all galaxies with very high S/N we
recover the expected ICD to within 0.1\%. However, as the S/N decreases
Equation~\ref{eqn:xi} systematically \textit{underestimates} the ICD. This bias is most
apparent for galaxies with ``low'' intrinsic ICD.

Because of the systematic underestimation of the ICD it is important to clip the sample at a
S/N which still provides accurate results. We choose S/N = 10 such that the bias is does not
impact our results. For example, the bias is the fractional change ($\xi_{true} -
\xi_{observed} / \xi_{true}$). For objects with S/N = 10 and high intrinsic ICD ($\xi_{true}
> 6\%$), the bias is small, with relative changes of $<34\%$, and the bias decreases for
objects with increasing ICD. The bias increases for objects with lower intrinsic ICD, such
that the relative change is 100\% for objects with S/N = 10 and $\xi_{true} = 2\%$. This is
acceptable because the ICD increases by only a factor of 2. Therefore, to our S/N = 10
limit, we can be confident that objects with observed high ICD have high intrinsic ICD, and
objects with low observed ICD have low intrinsic ICD.

In principle this S/N requirement applies to both bands used in Equation~\ref{eqn:xi}.
However, for this particular study, the redder band (namely \wfch) S/N is much higher in the
majority of galaxies analyzed, such that this requirement, in practice, only applies to the
bluer imaging band. It is also important to note that this S/N does not correspond to a
single bluer band magnitude limit, because at fixed magnitude a larger object (with lower
surface brightness) will have a lower S/N. For galaxies with an intrinsic ICD very near
zero, and low surface brightness in the bluer band, it is possible for the ICD calculation
to produce negative results when the background contributes significantly to the
fluctuations. Galaxies with an intrinsically larger ICD are more resistant to surface
brightness dimming as the fluctuations of the galaxy dominate those of the background.

After applying the S/N = 10 restriction in \acsi, we have a final, primary sample of 2601
galaxies. Unfortunately, this S/N requirement leads to a selection effect where we prefer
galaxies with some level of active star-formation, as passively evolving galaxies often do
not have enough blue light to be included in this analysis. The effect of this selection
bias is discussed further in Section~\ref{sec:icdvmass}.

\subsection{Error Associated with ICD}\label{sec:ICD error} \cite{Papovich2003} defines the
statistical error associated with the ICD, assuming that the background pixels (those not
associated with any galaxy) are normally distributed, as \begin{equation}\label{eq:ICD
error} \delta({\xi}) = \frac{\sqrt{2/N_{pix}} \sum\limits_{i=1}^{N_{pix}} (B_{2,i}^2 +
\alpha^2 B_{1,i}^2)} {\sum\limits_{i=1}^{N_{pix}} (I_{2,i} - \beta)^2 -
\sum\limits_{i=1}^{N_{pix}}(B_{2,i} - \alpha B_{1,i})^2}. \end{equation} \cite{Papovich2005}
note that Equation~\ref{eq:ICD error} underestimates the uncertainties associated with the
ICD. They estimated that the uncertainties for Equation~\ref{eqn:xi} in their data required
corrections of factors of 6-7 to account for correlated noise in the images. \cite{Bond2011}
find that even after the correction suggested by \cite{Papovich2005}, the uncertainty on the
ICD obtained for bright objects (those with low errors) are underestimated by as much as a
factor of three. The dominant source of error being contributions from non-uniform sky
backgrounds and PSF mismatch between the two passbands.

For this work, we use the observed ICD values for different levels of intrinsic ICD and
galaxy S/N to estimate the error associated with each measurement. To estimate the
systematic uncertainty we compare the median ICD fractional change (the dashed line in
Figure~\ref{fig:ston_I}) to the true value, and to estimate the statistical uncertainty we
compute the median absolute deviation.

In all cases, galaxies with very high (\acsi\ S/N $>100$) S/N have low ($<1\%$) systematic
uncertainties and very low ($\ll0.2\%$) statistical uncertainties. The closer to our \acsi
S/N cutoff (\acsi S/N = 10) the systematic error, described in the previous section,
increasingly dominates. The statistical uncertainty remains very low $\sim0.2\%$ and so we
neglect any effect due to the statistical variation in the ICD.

%%%%%%%%%%%%%%%%%%%%% %%% BEGIN RESULTS %%% %%%%%%%%%%%%%%%%%%%%%

\section{THE RELATION BETWEEN GALAXY PROPERTIES AND INTERNAL COLOR
DISPERSION}\label{sec:results} For the majority of our results we choose to use $\xi(\acsi,
\wfch)$ as our primary diagnostic. For the galaxies in our sample, these two filters span
the Balmer/4000 \AAA break over $z\sim1.5-3$, and allow us to compare the rest-frame UV
(traced by the \acsi\ band) to the rest-frame optical (traced by the \wfch\ band). This
provides us the ability to investigate recent star formation (with \acsi) and past-averaged
star formation (with \wfch). In several sections (Section~\ref{sec:icdvmass} and
Section~\ref{sec:redshift}) we make use of different versions of the ICD, namely $\xi(\acsv,
\wfcj)$ which allows us to compare slightly different parts of each galaxy's spectral energy
distribution to better understand the results given by $\xi(\acsi, \wfch)$.

The median ICD for the full sample is $\xi(\acsi, \wfch) \simeq 5\%$. In the sections that
follow we refer to galaxies with ICD value above and below this median as “high” and “low”
ICD, respectively

\subsection{The Internal Color Dispersion vs. Galaxy Mass}\label{sec:icdvmass}
\begin{figure*} \includegraphics[width=\textwidth]{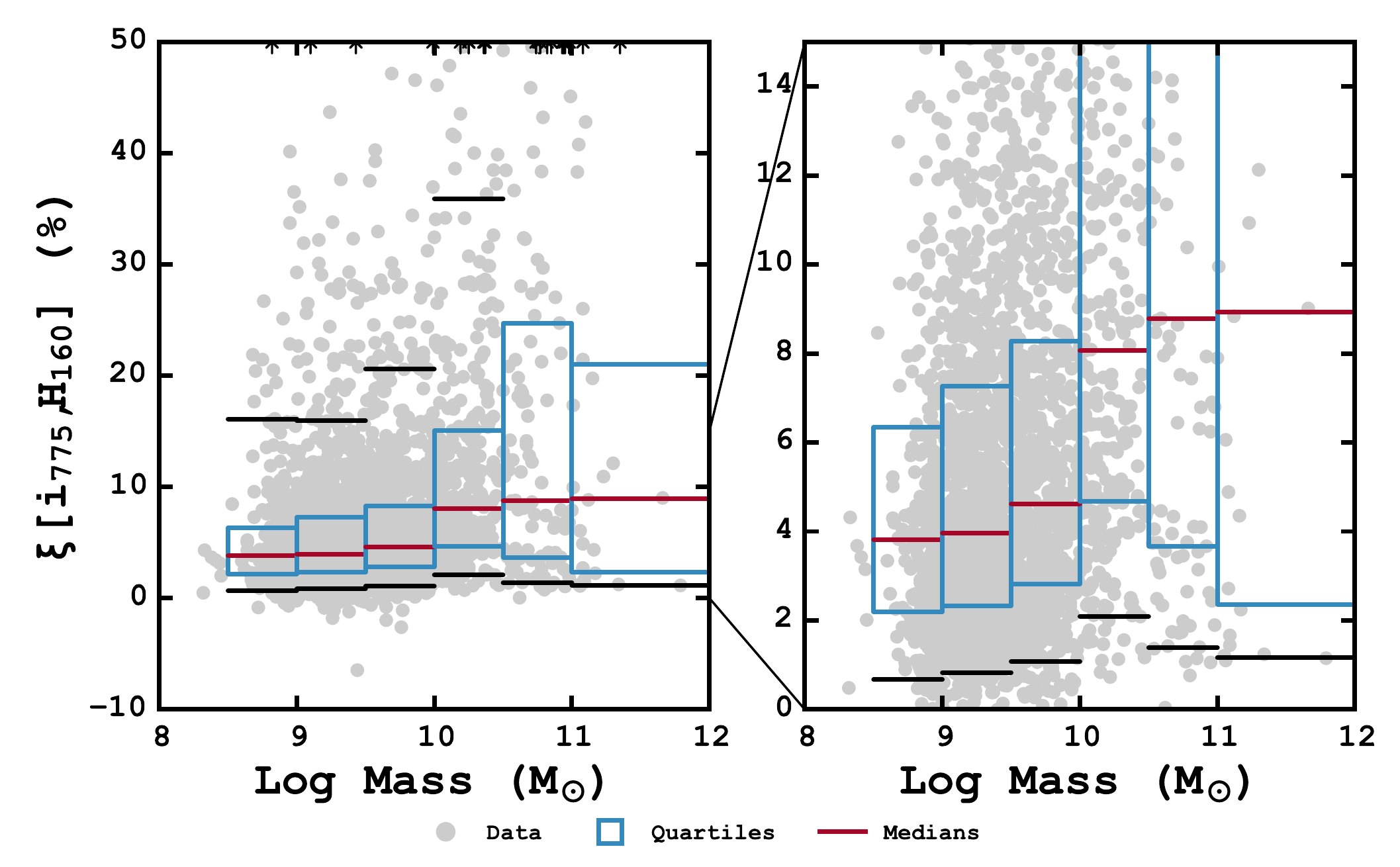} \caption{The ICD as a function of galaxy stellar mass. Individual galaxies are shown as gray points. The galaxies are then binned together by stellar mass in bins of 0.5 dex (except for the highest mass bin which is 1.0 dex) beginning with $\mathrm{Log~M}/\Msol =8.5$. The upper and lower quartiles of ICD are
indicated by the blue boxes. The median of each bin is indicated by the red line. The upper
and lower black bars show the 95th and 5th percentiles. The ICD for galaxies in our sample
is low for low stellar masses, peaks in the intermediate stellar masses, and returns to low
values at high stellar masses.} \label{fig:icd_vs_mass} \end{figure*}

Stellar mass is one of the most fundamental properties of galaxies. The stellar mass of a
galaxy correlates with many of its physical properties including overall color
\citep{Grutzbauch2011} and SFR (\eg, \citealt{Daddi2007a, Magdis2010a}) to $z\sim3$ (and
perhaps beyond \eg, \citealt{Salmon2014}). As such, it is possible to gain much insight into
the formation and evolution of galaxies through the rates of change of the ICD as functions
of stellar mass.

Figure~\ref{fig:icd_vs_mass} shows the $\xi(\acsi, \wfch)$ as a function of stellar mass for
the galaxies in our sample. We find the median $\xi(\acsi, \wfch)$ values in the three low
stellar mass bins are between $3.5\%-4.5\%$, the median ICD value in the moderate stellar
mass bin is above 8\%, and at stellar masses $\mathrm{Log~M} / \Msol > 11$, we find that
galaxies have a median ICD value about 9\%. If we take into account the systematic
underestimations of low ICD values, discussed in Section~\ref{sec:ICD error}, we find little
relative change in the median $\xi(\acsi, \wfch)$ between bins, in that the moderate stellar
mass bins still show higher $\xi(\acsi, \wfch)$ medians than the bins of lower stellar
masses.

\begin{figure*} \includegraphics[width=\textwidth]{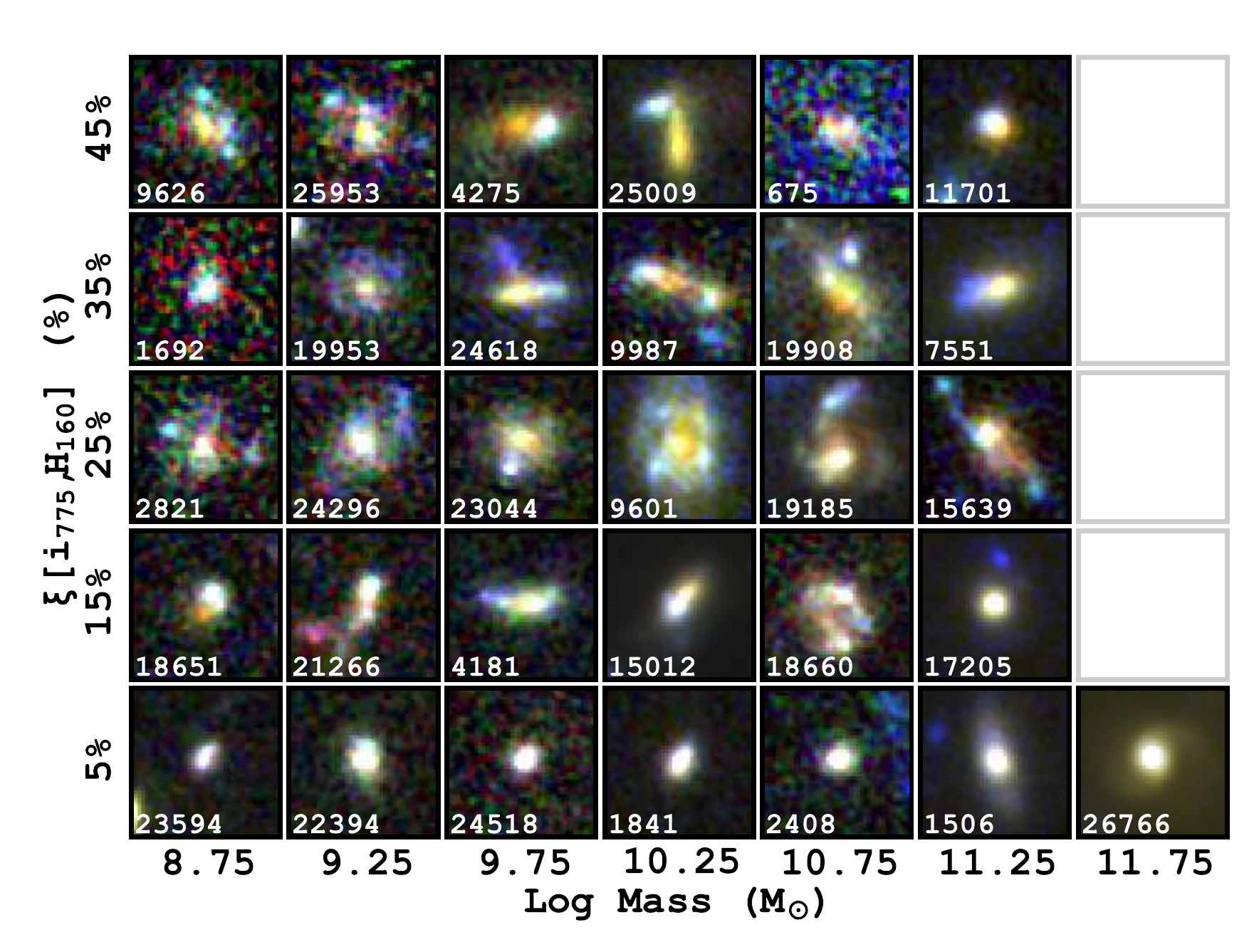} \caption{Galaxy montage showing examples of galaxies as a function of their location on the ICD - stellar-mass plane. Columns are bins of stellar mass while rows are bins of ICD. The RGB images are combinations of \wfch, \wfcj, and \acsi. Galaxies with elevated ICD values ($\xi(\acsi, \wfch) > 15\%$) and stellar masses $\mathrm{Log~M}/\Msol~10-11$ appear significantly different than the galaxies in either other stellar mass bin or galaxies with depressed ICD values. Each cutout is $2.5''$ on the sky regardless of the redshift of the object, which corresponds to $\sim20$ kpc over the redshift range. Columns are bins of stellar mass while rows are bins of ICD. Galaxies filling the grid are selected at random from all galaxies falling into the bin. Empty squares indicate regions in the $\xi(\acsi, \wfch)$-mass space where there are no galaxies in our sample. The labels represent the centers of either the ICD or stellar mass bins. So a galaxy in the bottom left corner (\eg\ 23594) could have a stellar mass $8.5 <
\mathrm{Log~M}/ \Msol < 9$ and and ICD of $0\% < \xi(\acsi, \wfch) < 10\%$. }
\label{fig:montage} \end{figure*}

To illustrate the ICD, Figure~\ref{fig:montage} shows galaxy images in the $\xi(\acsi,
\wfch)$-mass plane. The galaxies are selected at random from all of the galaxies contained
in each bin of mass and $\xi(\acsi, \wfch)$. The RGB images are combinations of \wfch\,
\wfcj, and \acsi. Galaxies with elevated ICD values (top rows) and in the moderate stellar
mass range appear significantly different than the galaxies in other stellar mass bin or
galaxies with lower ICD values. Visually, galaxies with high ICD have morphologies with
well-separated regions of rest-frame UV (blue) and rest-frame optical (red) light. Galaxies
with low ICD values tend to have smooth morphologies and uniform colors, suggesting fewer
regions of separated red and blue emission.

\subsubsection{High Mass Galaxies} The highest stellar mass galaxies ($\mathrm{Log~M}/\Msol
>11$) deserve special attention. Because the S/N threshold applies only to the bluer band
used in the ICD calculation, galaxies which are well detected in \wfch\ may not be well
detected in \acsi. Specifically, massive galaxies often have the reddest $\acsi - \wfch$
colors, so we expect that many of them have high \wfch\ S/N and low \acsi\ S/N. The concern
is that these objects could have higher intrinsic ICD, but are excluded because of our
signal to noise requirement (see Section~\ref{sec:ICD vs ston}).

To investigate this possibility, we compare the mean $\acsi - \wfch$ color of the high
stellar mass galaxies included and excluded from our sample. Galaxies which have
S/N$_{\acsi} \geq 10$ have $\langle\acsi - \wfch\rangle = 2.6$ mag while galaxies with
S/N$_{\acsi} \leq 10$ have $\langle\acsi - \wfch\rangle = 3.4$ mag. This suggests that the
excluded high stellar mass galaxies simply do not have enough blue light to produce high ICD
values. \cite{Papovich2005} simulate just such a situation where the authors find that
extremely red galaxies, even with the addition of highly star-forming regions produce nearly
no ICD as the older stellar populations simply overwhelm any diversity. For the seven high
stellar mass galaxies with $\acsi - \wfch > 3$ mag in our sample, four have high $\xi(\acsi,
\wfch)$; however, a visual inspection provides additional insight into these objects
(discussed below).

\begin{figure*} \includegraphics[width=\textwidth]{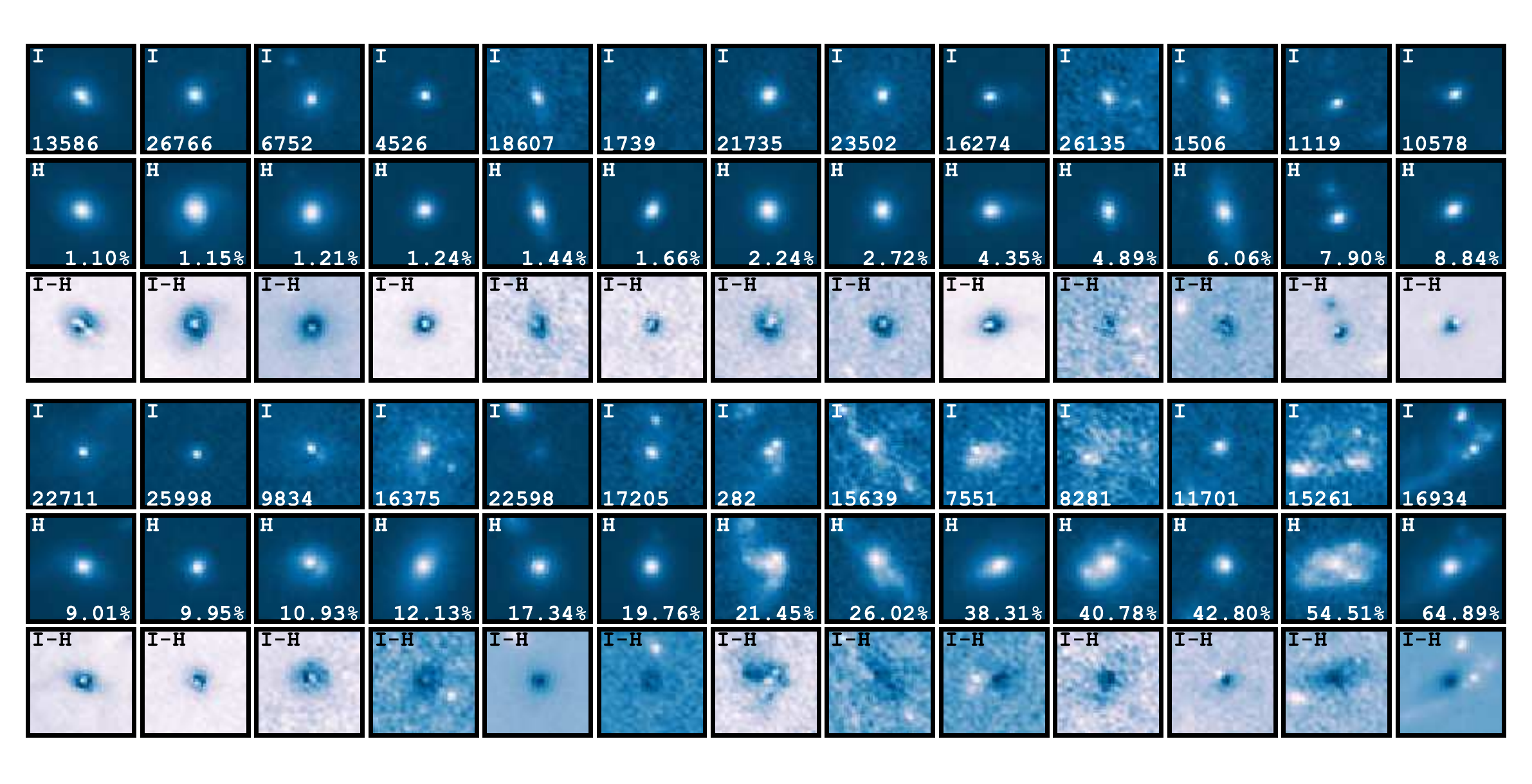} \caption{Galaxy montage of the high mass (11 $>\mathrm{Log~M}/\Msol$) galaxies. Each triplet of panels shows either the \acsi, \wfch\ imaging or the $\acsi - \wfch$ color map, along with the ID number and associated $\xi(\acsi, \wfch)$ value. Each cutout is $2.5''$ on the sky regardless of the redshift of the object, which corresponds to $\sim20$ kpc over the redshift range. The panels are ordered from lowest to highest $\xi(\acsi, \wfch)$.} \label{fig:highmass_stamps}
\end{figure*}

A visual inspection of the 26 most massive galaxies, shown in Figure~\ref{fig:highmass_stamps} shows that the majority of these objects have smooth light
profiles in both \acsi, and \wfch. There are several exceptions which fall into two broad
categories: galaxies which appear to have faint, nearby objects detected in \acsi\ but not
\wfch\ (\eg, 16375, 17205 and 16934). Because \wfch\ serves as the object detection image,
it is possible that these faint objects lie within the \wfch\ pertrosian radius used for the
ICD calculation, contributing significantly to the measured ICD. While the origin of these
nearby objects is unclear, possible scenarios include a chance alignment or a disrupted
satellite. And, galaxies in the highest-mass subsample with extended structures, showing
distinct red/blue regions and likely individual “clumps” of star-formation embedded within
the galactic structures (\eg, 282, 15639 and 15261). In both cases, these features contrast
with the \wfch\ dominated central region producing the large $\xi(\acsi, \wfch)$ value. The
blue central regions in the $\acsi - \wfch$ color maps reflects small mismatches in the PSF
matching (see \citealt{VanderWel2012} for a description) and do not contribute significantly
to the ICD as they are small compared to the total flux of the galaxy.

After a visual inspection, if we remove the galaxies with neighboring objects from the high
stellar mass bin the median $\xi(\acsi, \wfch)$ becomes $\sim7\%$. However, if we instead
place a more stringent \acsi\ S/N requirement (\acsi\ S/N $>30$) on the high stellar mass
galaxies, we find the median $\xi(\acsi, \wfch)$ further drops to 3.3\%.

Because it is unclear whether or not the faint objects neighboring the primary galaxy are
physically associated with the primary, or are chance alignments along the line of sight, we
do not exclude these objects and do not adjust the \acsi\ S/N requirement. Instead, we note
that these are objects are infrequent in the sample, and so do not affect our conclusions.
So while the median $\xi(\acsi, \wfch)$ in the high stellar mass bin may appear to be very
similar to the moderate stellar mass bins, it is likely that the median is being inflated
due to inclusion of galaxies with neighboring objects or galaxies with low \acsi\ S/N and
high \wfch\ S/N.

\subsection{The Internal Color Dispersion vs. S\'{e}rsic Profile}\label{sec:sersic} Many
previous works (\eg, \citealt{Blanton2003a, Bell2012, VanderWel2012}) use the S\'{e}rsic
index to classify a galaxy's morphology. The S\'{e}rsic index parameterizes a galaxy's
surface brightness profile, where $n=4$ corresponds to a classical bulge, $r^{1/4}$-law
profile \citep{DeVaucouleurs1948} and $n=1$ corresponds to the exponential profile of a
galactic disk. We refer to galaxies with $n<2$ as disk-dominated, and those with $n>2$ as
bulge-dominated.

\begin{figure*} \includegraphics[width=\textwidth]{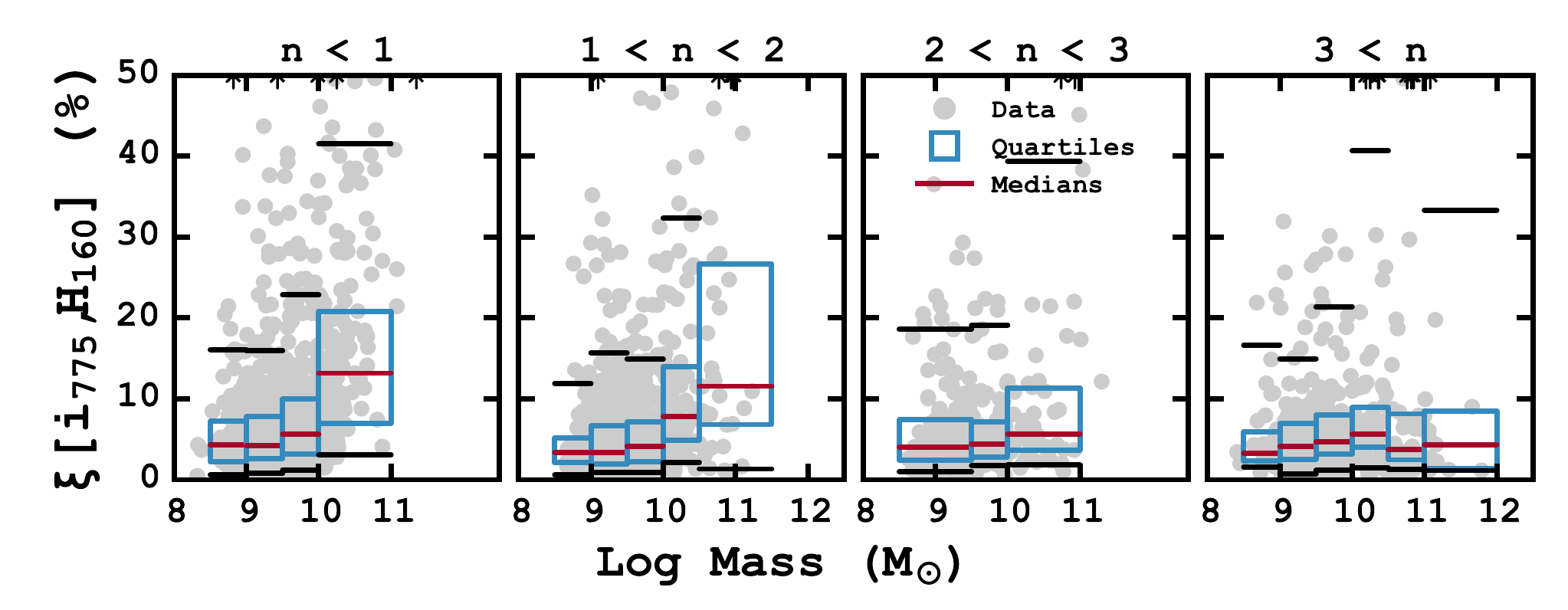} \caption{Galaxies binned into four bins of S\'{e}rsic index. The galaxies are then binned by stellar mass in bins of 0.5 dex beginning with Log $\mathrm{M}/\msol = 8.5$. The upper and lower quartiles are indicated by the blue boxes. The medians of each bin is indicated by the red line. The upper and lower bars show the 95th and 5th percentiles. We find that galaxies with high ICD (greater than 5\%) are dominated by low S\'{e}rsic indices in general suggesting they are either disk galaxies or have high extinction.} \label{fig:icd_vs_mass_vs_sersic} \end{figure*}

Figure~\ref{fig:icd_vs_mass_vs_sersic} shows the $\xi(\acsi, \wfch)$ as a function of
stellar mass, separated based on their S\'{e}rsic indices. Again, we divide the sample into
mass bins, where each bin is 0.5 dex wide or larger to ensure there are at least 10 objects
in each bin. Galaxies with high ICD prefer a lower S\'{e}rsic index (disk dominated). Inside
of the individual S\'{e}rsic bins, galaxies with intermediate stellar mass ($10 <
\mathrm{Log~M}/ \Msol < 11$) display the largest ICD, but as the S\'{e}rsic index increases,
the upper envelope of ICD values decreases: bulge-dominated ($n > 2$) galaxies have low ICD.
For disk-dominated galaxies ($n<2$) the median galaxy across all masses has a high ICD value.

The systematic underestimation described in Section~\ref{sec:ICD error}, has very little
effect on the preference of galaxies with low S\'{e}rsic index and intermediate stellar mass
to have high $\xi(\acsi, \wfch)$. This is because there is very little relative change in
the medians of the bins due to the underestimation of the ICD.

There is a well known trend between the S\'{e}rsic index and stellar mass, with the
S\'{e}rsic index increasing with increasing stellar mass, This relationship is apparent both
at low (\eg, \citealt{Dutton2009}) and high (\eg, \citealt{Franx2008, Bell2012}) redshift.
We find a similar trend. The three low S\'{e}rsic bins lack (or have very few) galaxies with
masses greater than $\mathrm{Log~M}/ \Msol \sim 11$. In fact, it is only the highest
S\'{e}rsic bin which shows any significant number high stellar mass galaxies.

\subsection{The Internal Color Dispersion vs. Star Formation Rate}\label{sec:icdvssfr} 
\begin{figure} \includegraphics[width=0.49\textwidth]{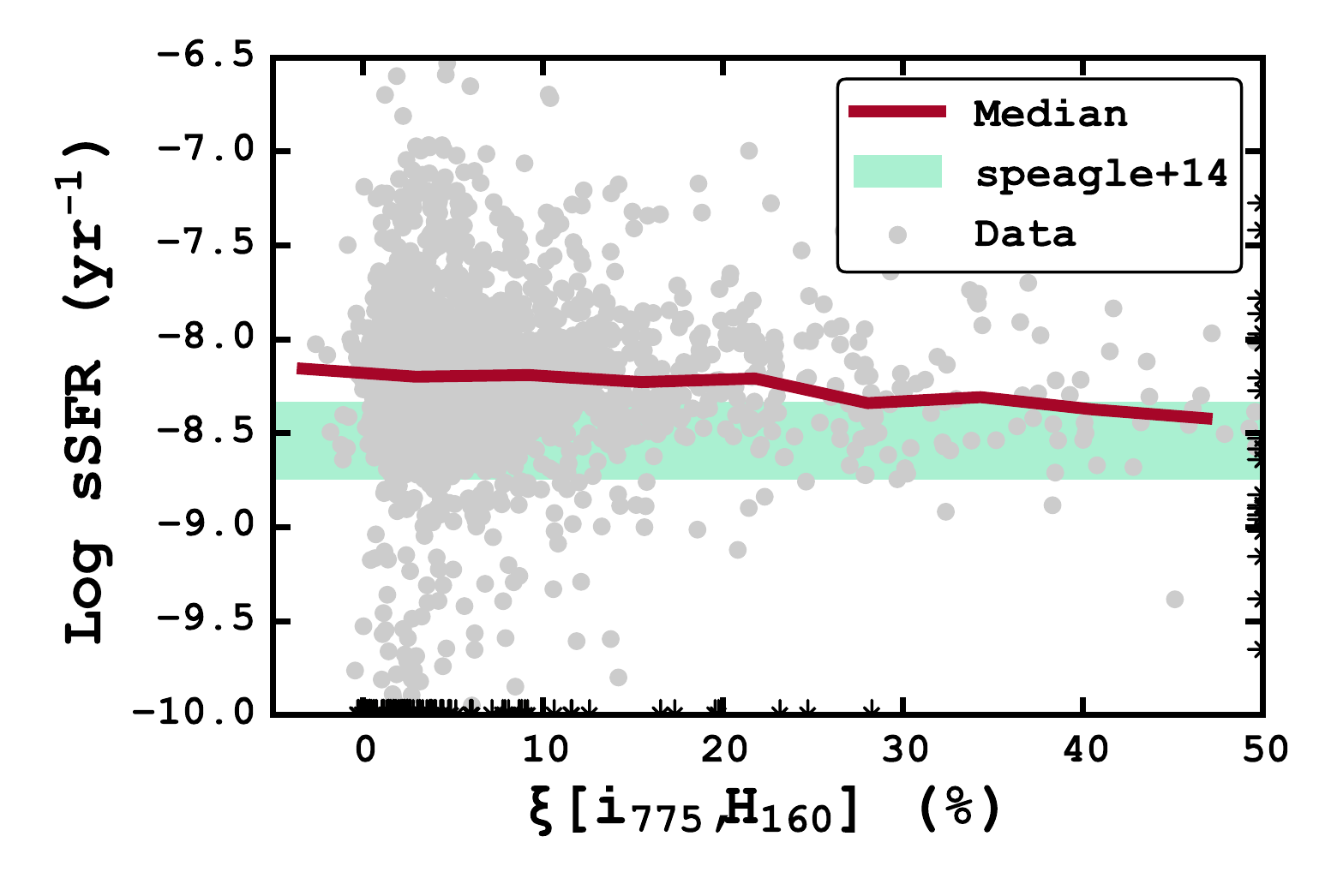} \caption{Galaxies' specific star formation rate (sSFR) as a function of $\xi(\acsi, \wfch)$. The gray points are the actual data values and the red line is the running median value. Galaxies which have been clipped for clarity are shown as black arrows. The shaded region shows the width of the star-forming ``main sequence'' for a $10^{10}\msol$ stellar mass galaxy at $z=2.25$ taken from \cite{Speagle2014a}. We find that galaxies with the highest $\xi(\acsi, \wfch)$ have very close to the median sSFR while galaxies with the highest and lowest sSFRs lack many galaxies with very high $\xi(\acsi, \wfch)$ values.} \label{fig:icd_sfr} \end{figure}

The
ICD is sensitive to heterogeneities in the stellar populations of galaxies. One possible
cause of these heterogeneities is unobscured, young stellar populations separated from
older, passively evolving populations. Because young populations are needed to produce
measurable ICD, the star formation rate (SFR) of a galaxy could be closely related to the
ICD. Figure~\ref{fig:icd_sfr} shows the specific star formation rate (sSFR), the star
formation rate normalized by the galaxy's stellar mass, as a function of $\xi(\acsi, \wfch)$.

We find that the $\xi(\acsi, \wfch)$ of a galaxy in not directly tied to its sSFR. Galaxies
with low $\xi(\acsi, \wfch)$ values show no correlation with sSFR, however, galaxies with
high $\xi(\acsi, \wfch)$ values show predominately ``normal'' sSFRs. For galaxies with low
sSFRs we expect low $\xi(\acsi, \wfch)$ values as such galaxies have few young stellar
populations producing the blue light needed to contrast the redder light from more evolved
stellar populations. Galaxies showing very high sSFRs, often due to re-radiation by dust in
the IR, also have low $\xi(\acsi, \wfch)$ suggesting that dust variations play a small role
in the production of ICD.

Again, this result is robust even after accounting for the systematic underestimate of the
low $\xi(\acsi, \wfch)$ galaxies. Shifting the lowest $\xi(\acsi, \wfch)$ galaxies upward to
higher $\xi(\acsi, \wfch)$ values does not produce a trend with sSFR.

%%%%%%%%%%%%%%%%%%% %%% END RESULTS %%% %%%%%%%%%%%%%%%%%%%

\section{POTENTIAL DRIVERS OF THE ICD}\label{sec:high icd} As discussed previously, the ICD
is sensitive to heterogeneous stellar population variations that give rise to differences in
the UV-optical flux ratio within a galaxy. In this section we discuss possible scenarios
which could lead to high ICD. Specifically, we investigate whether merger activity, the
galaxy's physical size, intrinsic color gradients, effects due to redshift, or heterogeneous
extinction drive increased ICD.

\subsection{Mergers}\label{sec:mergers} Based on morphological asymmetries,
\cite{Papovich2005} suggested that some of the galaxies with high ICD are likely the result
of merger activity. To investigate this, we use non-parametric morphological indicators,
specifically the Gini coefficient, $G$, and the second order moment of the brightest 20\% of
a galaxy's flux, $M_{20}$, to better understand the role of galaxy mergers in producing high
ICD. Many studies demonstrate these parameter's usefulness in quantifying galaxy morphology
in large samples at both low and high redshifts (\eg, \citealt{Lotz2004a, Abraham2007,
Law2007, Law2012, Wang2012, Lee2013}).

\begin{figure} \includegraphics[width=0.49\textwidth]{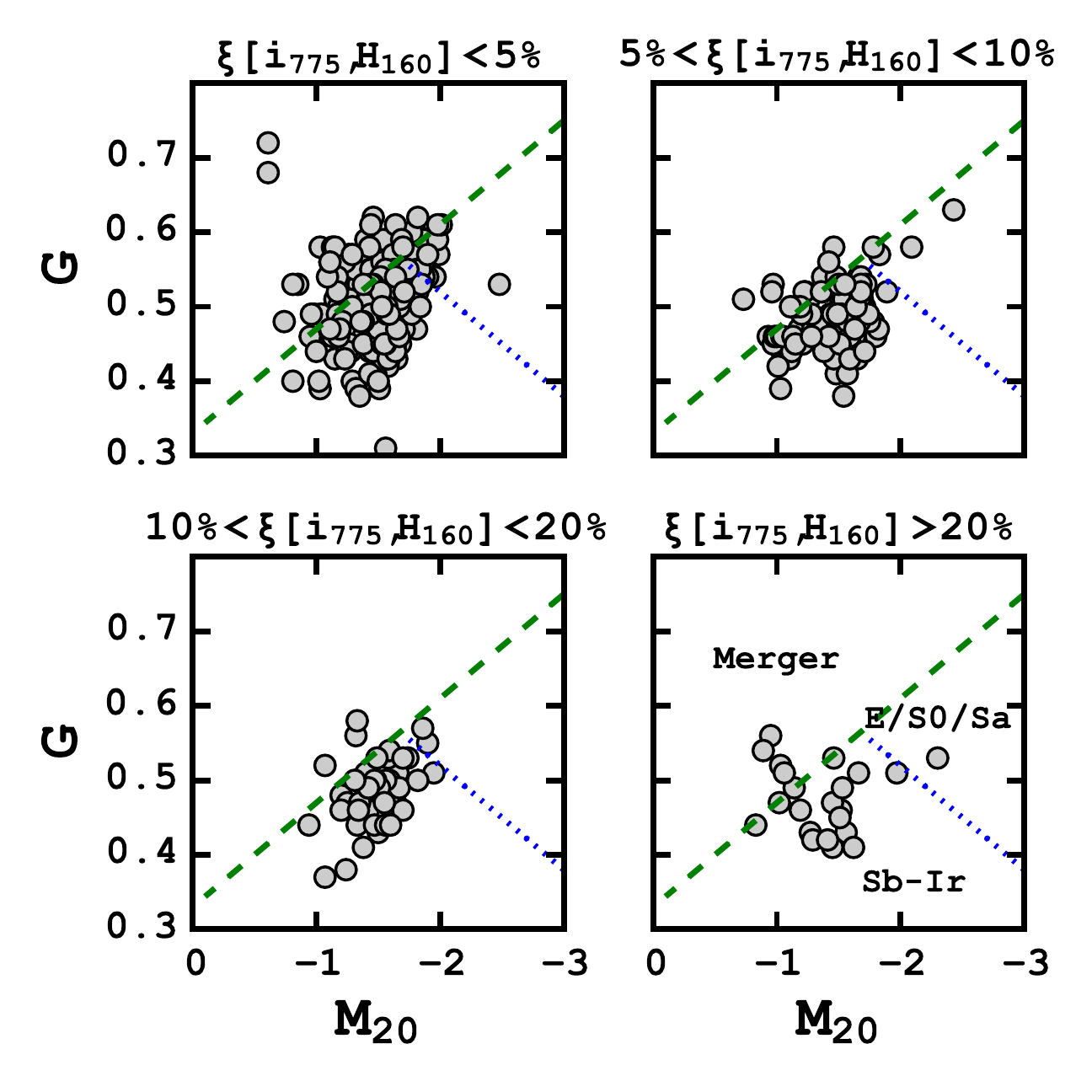} \caption{Morphological classification ($G-M_{20}$) of objects in our sample. Based on this classification scheme, objects above the dashed, green line are mergers, below the dashed, green line but above the blue, dotted line are elliptical galaxies and below both lines are spiral galaxies. The majority of galaxies with $\xi(\acsi, \wfch) > 5\%$ are not classified as mergers.} \label{fig:gini_m20} \end{figure}

\cite{Lotz2008} adopt the following classification scheme based on G-$M_{20}$:
\begin{eqnarray} \mathrm{Mergers:}~G > -0.14 M_{20} + 0.33 & & \\ \mathrm{E/S0/Sa:}~G \leq
-0.14 M_{20} +0.33 &~\mathrm{and}~& G> 0.14 M_{20} + 0.80 \nonumber \\ \mathrm{Sb - Ir:}~G
\leq -0.14 M_{20} + 0.33 &~\mathrm{and}~& G \leq 0.14 M_{20} + 0.80 \nonumber \end{eqnarray}

Figure~\ref{fig:gini_m20} shows the $G$ and $M_{20}$ values for the galaxies in our sample
broken into large bins of ICD. As can be seen, mergers are not a dominant driver of the ICD.
The majority of galaxies with high ICD values ($5\% - 20\%$) prefer disk-dominated
morphologies, consistent with our findings between the ICD and S\'{e}ric index above. This
suggests that mergers are not a significant cause of elevated ICD, except, possibly, at the
highest levels ($\xi(\acsi, \wfch) > 20\%$) where 21\% of galaxies are classified as mergers.

Given the small number of galaxies in the highest $\xi(\acsi, \wfch)$ bin, we use a series
of Monte Carlo simulations to estimate how often we would find a similar distribution simply
by chance. We assign each galaxy a morphology based on their location in the $G-M_{20}$
space while allowing their $\xi(\acsi, \wfch)$ to be changed. We then select galaxies with
$\xi(\acsi, \wfch) > 20\%$ and record the fraction classified as mergers. We find that
galaxies in the highest ICD bin are at least 21\% mergers only 1.32\% of the time. Or, given
just random chance, 98.7\% of the time we find fewer mergers in the highest ICD bin than
what is actually measured.

\subsection{Physical Size}\label{sec:size} The ICD requires that we are able to resolve
scales to a minimum of 0.5 kpc \citep{Papovich2003}. Below this, the resolution of the image
suppresses any ICD that may be present. Here we also considered how the ICD values change
with increasing physical size.

\begin{figure} \includegraphics[width=0.49\textwidth]{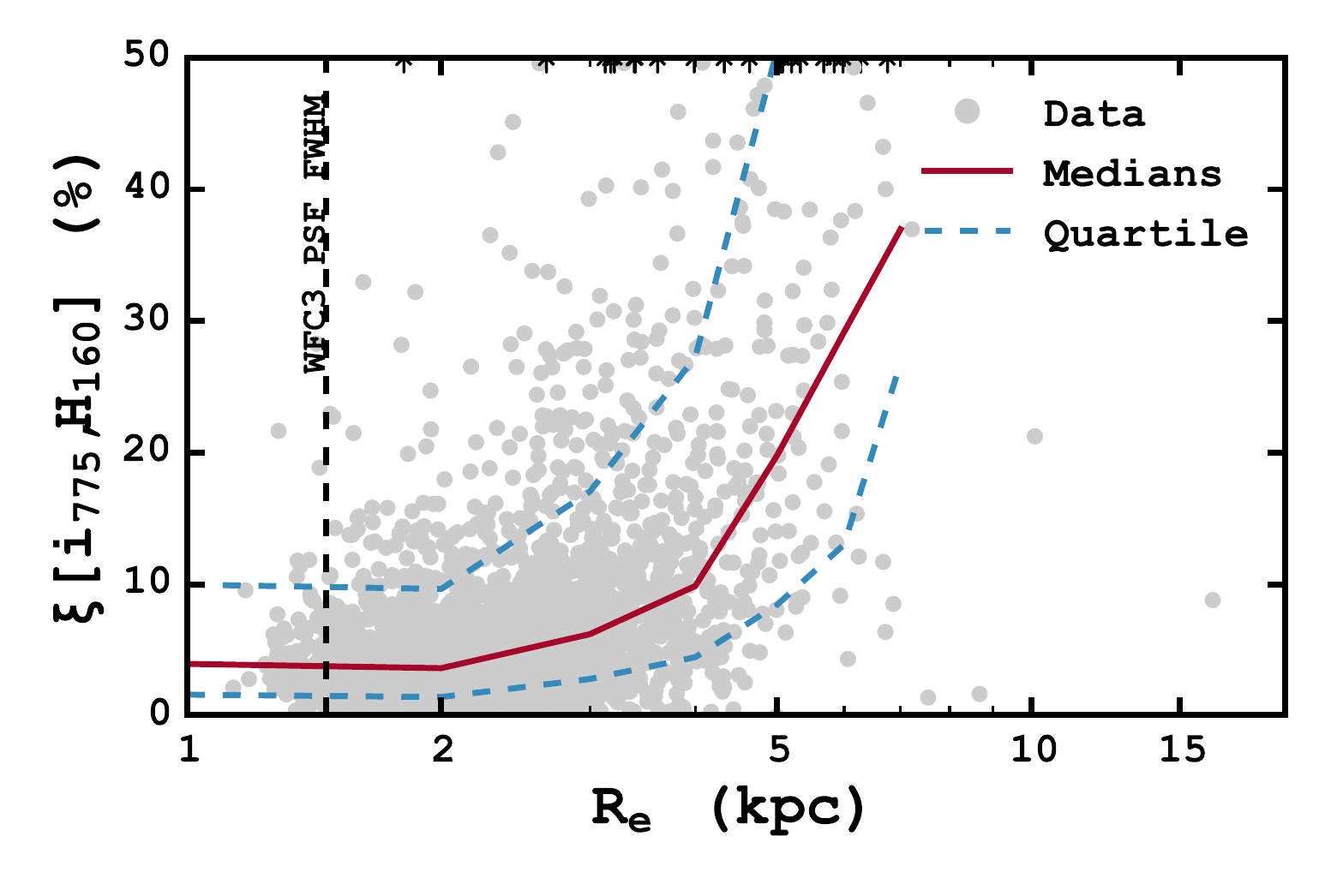} \caption{$\xi(\acsi, \wfch)$ as a function of the galaxy's effective radii (semi-major axis). Gray points are the individual galaxies. The red line is the median $\xi(\acsi, \wfch)$ value and the blue dashed represent the interquartile range, 75\% and 25\% respectively. The dashed vertical line shows the effective radius corresponding to the FWHM of the WFC3 PSF (FWHM = $0.2''$) at $z=3.5$. We find no trend suggesting a preference of physical size for a galaxy with elevated ($\sim 7\%$) $\xi(\acsi, \wfch)$.} \label{fig:halflight} \end{figure}

Figure~\ref{fig:halflight} shows $\xi(\acsi, \wfch)$ as a function of semi-major axis,
$R_e$. For galaxies with $R_e$ below 3 kpc the median ICD remains low ($\sim 5\%$).
Considering only the intermediate mass ($10 < \mathrm{Log~M} / \Msol < 11$) we find ICD
values remain low until $\sim2.5$ kpc after which it increases. In either case, we find more
galaxies with large ICD at large $R_e$. However, this is most likely a selection effect
where we are more able to detect ICD in extended objects. Furthermore, placing the galaxies
on a size--mass relation provides a similar result. At a fixed stellar mass, we find more
elevated $\xi(\acsi, \wfch)$ galaxies with larger radii than at small radii.

One of our findings is that the ICD values are low for lower stellar mass galaxies,
$\mathrm{Log~M} / \Msol < 10$. Because these galaxies have lower $R_e$ compared to higher
mass galaxies (\eg, \citealt{Franx2008, Williams2010}) we test if the trend between ICD and
$R_e$ affects this finding. We remove all objects with $R_e < 2$ kpc (at all stellar masses)
and recompute how the median ICD changes as a function of mass. The median $\xi(\acsi,
\wfch)$ for $\mathrm{Log~M} /\Msol < 10$ remains low, with $\xi(\acsi, \wfch)_{median}
\simeq 5\%$. The median $\xi(\acsi, \wfch)$ for moderate mass galaxies remains elevated,
$\xi(\acsi, \wfch)_{median} > 10\%$, for $10 < \mathrm{Log~M} / \Msol < 11$, and the
$\xi(\acsi, \wfch)$ declines at the highest masses. Therefore, the relation between ICD and
$R_e$ does not dominate our conclusions between ICD and stellar mass.

\subsection{Color Gradients} \label{sec:Color Gradients} Galaxies show a radial color
gradient when the color of the light in an annulus on the outskirts differs from the color
inside a central region (see Section~\ref{sec:data_colorgraident}). For galaxies with large
color gradients, this difference could potentially lead to large ICD values.
\cite{Szomoru2011} found that the majority of a mass-selected sample from the HUDF show
negative color gradients (redder cores and bluer outer regions). We find a similar result
for the galaxies in our sample, with an average color gradient $\Delta(\acsi-\wfch) = -0.12$
and 66\% of the sample showing a negative color gradient.

\begin{figure} \includegraphics[width=0.49\textwidth]{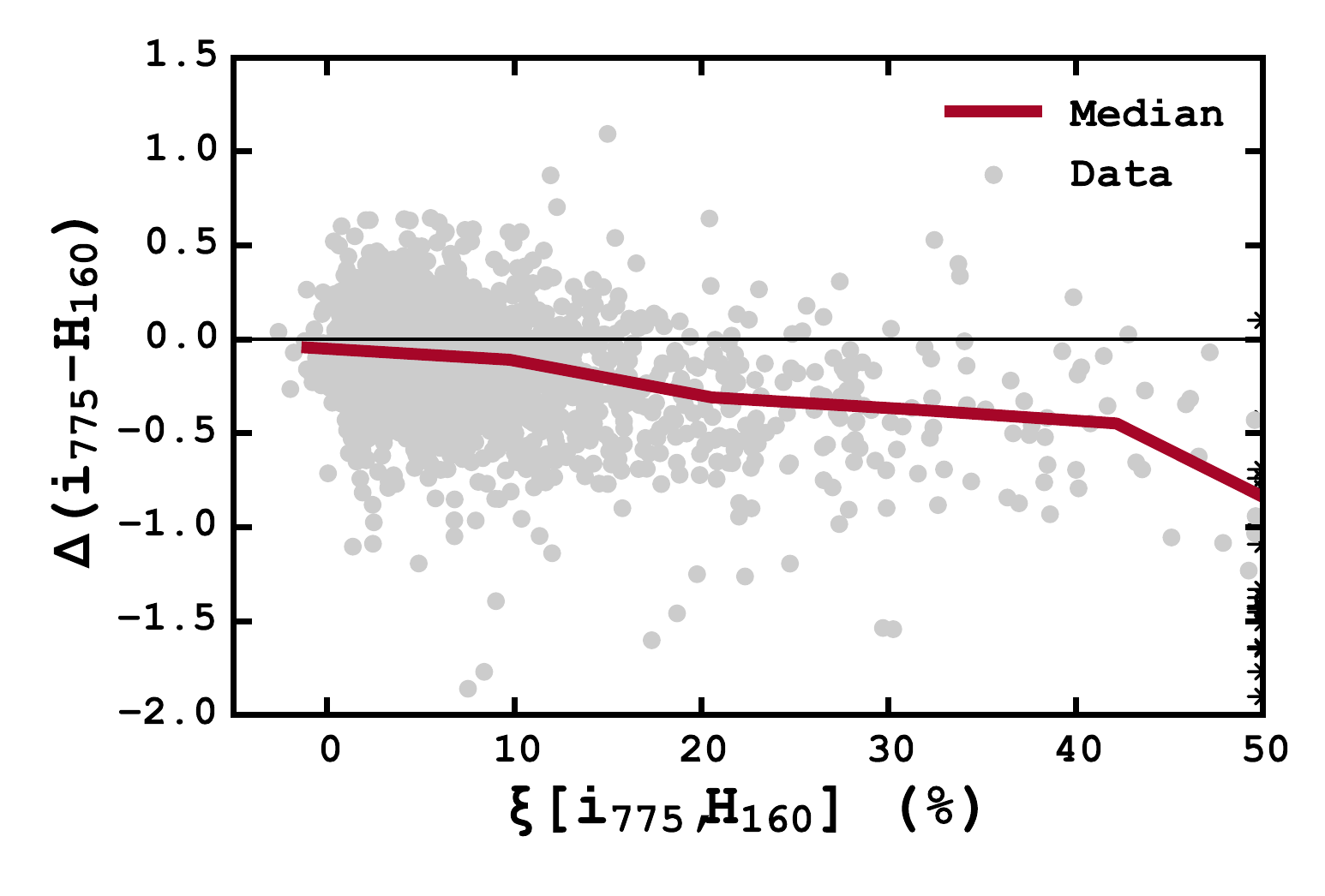} \caption{The galaxy's color gradient, $\Delta(\acsi - \wfch)$, as a function of $\xi(\acsi, \wfch)$. Gray points are the data and the red line is the median color gradient as a function of ICD. We find that galaxies with high ICD have systematically lower color gradients, suggesting they have redder cores than outskirts.} \label{fig:icd_vs_color_grad} \end{figure}

The relationship between color gradient and the ICD is complex.
Figure~\ref{fig:icd_vs_color_grad} shows $\Delta(\acsi - \wfch)$ as a function of
$\xi(\acsi, \wfch)$. We find that while galaxies with high ICD typically have negative color
gradients, there is no correlation between the strength of the color gradient and the ICD.
Therefore, while strong color gradients do not necessarily imply high ICD values, galaxies
with high ICD appear to require larger color gradients, either positive or negative.

\begin{figure} \includegraphics[width=0.49\textwidth]{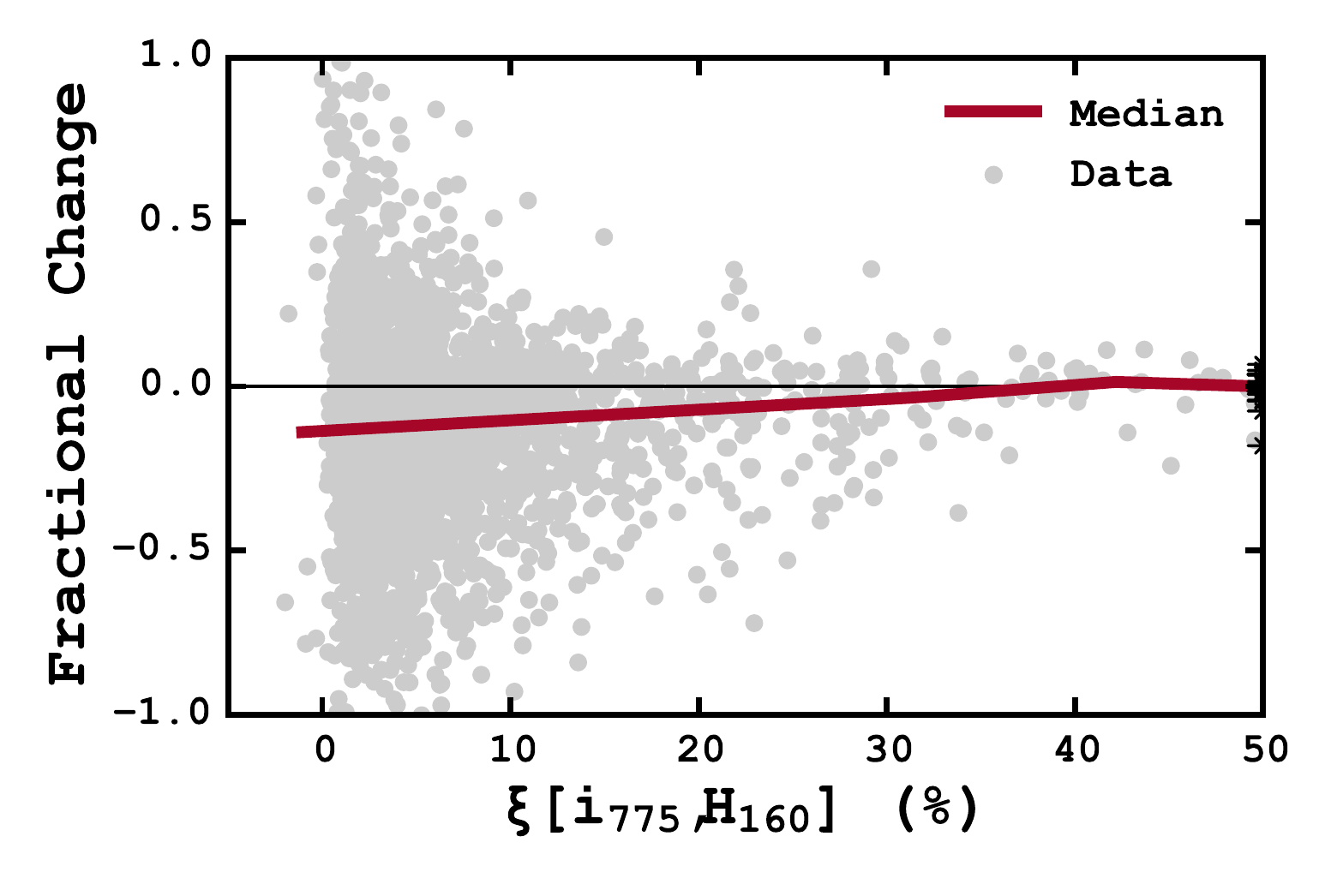} \caption{The fractional change between the $\xi(\acsi, \wfch)$ calculated with the center $0.18''$ removed and the $\xi(\acsi, \wfch)$ calculated for the entire galaxy as a function of $\xi(\acsi, \wfch)$ for the entire galaxy. We find that as galaxies with their centers removed show an elevated amount of $\xi(\acsi, \wfch)$. This is expected as the variations in the outskirts play a more significant role with the bright central region removed, and indicates the observed ICD is not due to an underlying color gradient.} \label{fig:icd_icdcored} \end{figure}

To understand how much a color gradient contributes to the measured ICD, we perform the
following test. We mask the central $0.18''$ ($\sim1.5$ kpc) of all galaxies and recompute
the ICD value. Figure~\ref{fig:icd_icdcored} shows the relative change in the ICD due to
this masking. In the majority of galaxies with small $\xi(\acsi, \wfch)$ values, removing
the nuclear region causes the ICD to decrease, while in galaxies with larger ICD values
there is little change. This is as expected as galaxies with small ICD values, masking out
the central regions also removes some of the variation. Therefore, while galaxies with
strong color gradients may have high ICD, this is not a requirement. The observed color
gradients alone are insufficient to produce the ICD in the galaxies in our sample.

\subsection{Redshift Effects from Bandpass Shifting on the ICD}\label{sec:redshift}
\begin{figure} \includegraphics[width=0.49\textwidth]{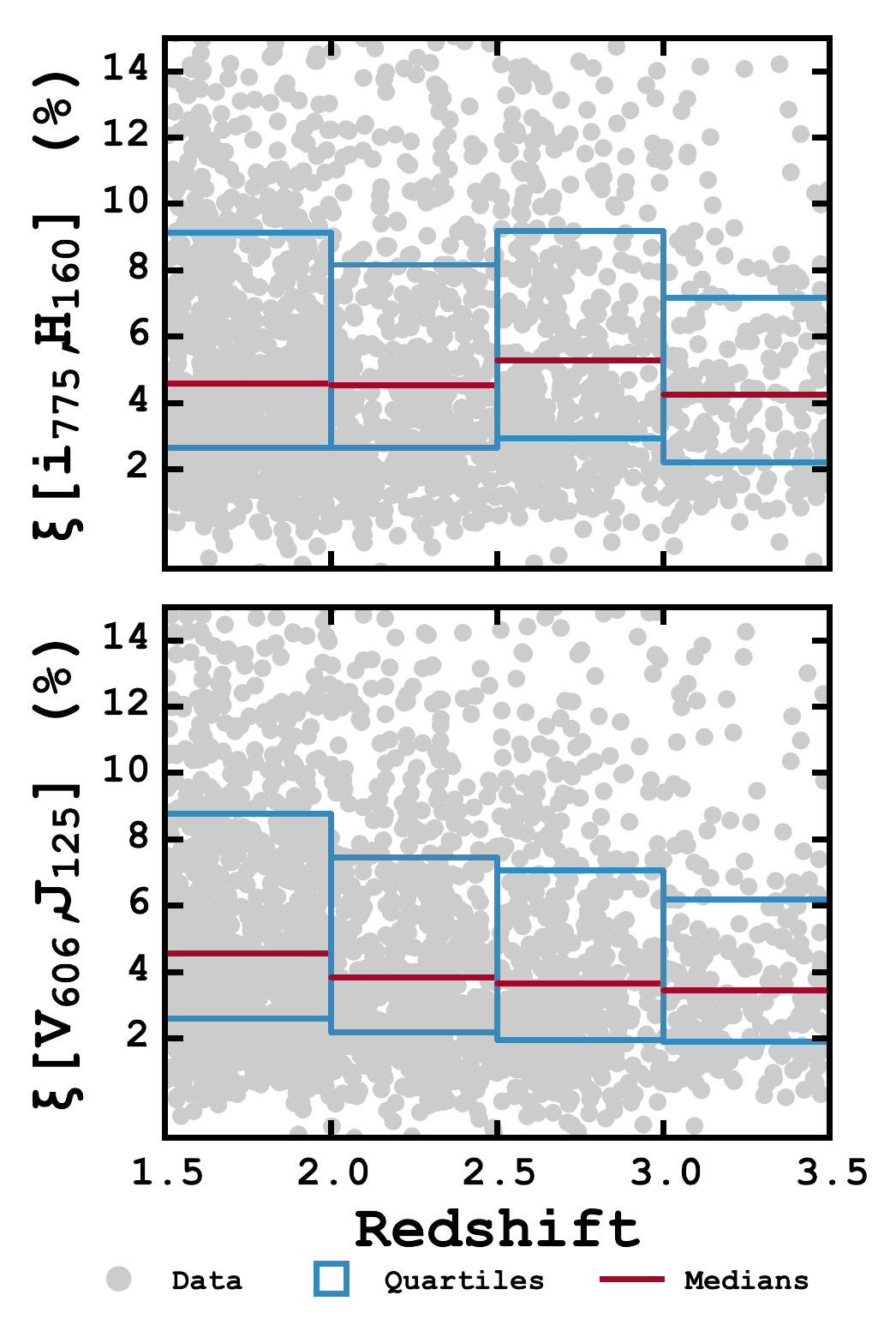} \caption{ICD of galaxies in our sample as a function of redshift. The galaxies are then binned together by redshift in bins of 0.5 dex beginning with $ z = 1.5$. The upper and lower quartiles are indicated by the blue boxes. The medians of each bin is indicated by the red line. For clarity, only a portion of the possible ICD range is shown. Top and bottom panel show $\xi(\acsi, \wfch)$ and $\xi(\acsv, \wfcj)$ as a function of redshift, respectively. In both panels we do not see a significant increase in the median values of ICD. The inclusion of the bluer bands (\acsv\ and \wfcj) acts as a probe for bandpass shifting effects. Because neither shows a significant trend, this suggests that the lack of increased values is not simply due to bandpass shifting effects.} \label{fig:icdvz} \end{figure}

The galaxies in our sample span a range of redshift from $1.5 < z < 3.5$. As a result, the
rest-frame light probed by both \acsi\ and \wfch\ varies between galaxies. For example, in
galaxies at $z=1.5-2.5$, \wfch\ probes rest-frame $4600-6400$ \AAA, while at higher
redshifts, $z>3$, \wfch\ probes light shortward of 4000 \AAA. Therefore, we have considered
how the variation in rest-frame light could affect our ICD measurement.

For galaxies at $z< 2$, we measure the ICD between \acsv\ and \wfcj, as these bandpasses
observe similar rest-frame wavelengths as \acsi\ and \wfch\ for galaxies at $z>2.2$.
Figure~\ref{fig:icdvz} shows ICD as a function of redshift where we have binned the galaxies
by redshift. The top row of Figure~\ref{fig:icdvz} shows $\xi(\acsi, \wfch)$ and the bottom
shows $\xi(\acsv, \wfcj)$. For the comparable bins ($z=1.5-2$ for $\xi(\acsv, \wfcj)$ and
$z\sim2.2$ for $\xi(\acsi, \wfch)$) we find very little ($<0.1\%$) difference between the
median ICD values. This shows that as long as the imaging bands chosen to calculate the ICD
span the rest frame Balmer/4000\AAA break we calculate a similar ICD.

For galaxies above $z=3$ both \acsi\ and \wfch\ lie blueward of the rest frame Balmer/4000
\AAA break, and \cite{Bond2011} suggests that a sharp downturn would be observed as the ICD
probes similar populations. We find no such downturn, suggesting that while the Balmer/4000
\AAA break is moving through \wfch, the observed trends in $\xi(\acsi, \wfch)$ as a function
of stellar mass (Figure~\ref{fig:icd_vs_mass}) is not simply due to the effect of redshift,
but due to the individual characteristics of the galaxies.

\subsection{Effects of Dust Attenuation on the ICD}\label{sec:attenuation} As mentioned
earlier, high ICD could result from spatial variations of stellar ages or dust attenuation.
In this section, we investigate whether the ICD is related to the total dust content of our
galaxies.

\begin{figure} \includegraphics[width=0.49\textwidth]{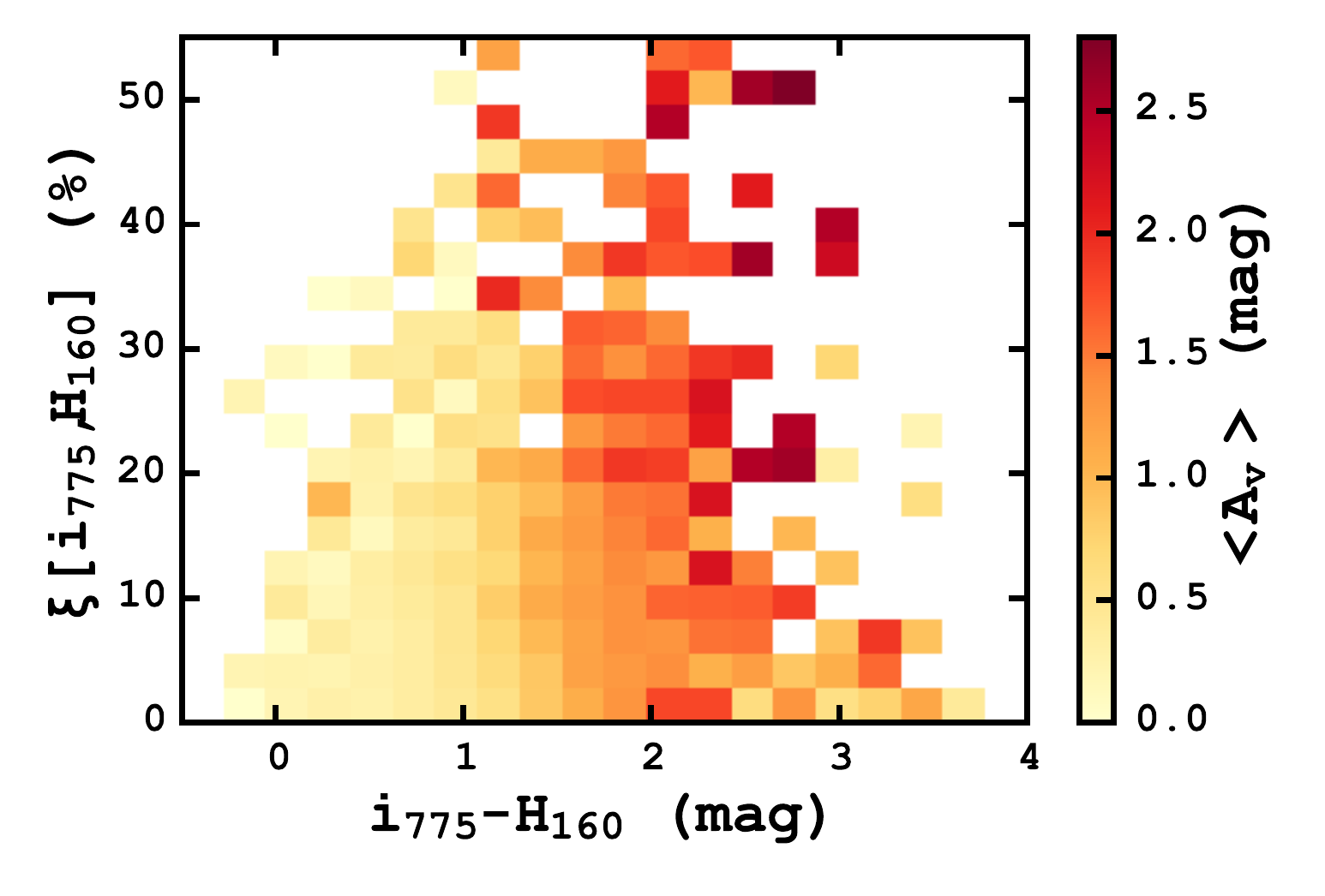} \caption{The Internal Color Dispersion as a function of $\acsi-\wfch$ color. The colored boxes show the average attenuation value ($\rm A_V$) of the points within the region. For the galaxies in our sample, we find no evidence of a correlation between the measured stellar population diversity and the rest frame UV-optical color. We do find that galaxies with large ICD values also show large attenuation values.} \label{fig:icd_vs_color} \end{figure}

If inhomogeneous attenuation (the reddening of selective parts of a uniformly blue
population) is the primary driver of the ICD then we would expect to find galaxies with both
very red colors and high ICD values. A uniformly blue galaxy would begin with a blue
$\acsi-\wfch$ color and if we were to add a level of inhomogeneous dust then we would expect
the ICD to rise, as there is now red light with which to contrast the blue and the overall
color to shift toward the red. Similarly if a galaxy with a heterogeneous stellar
populations has the young star forming regions attenuated then the ICD would go down as
there is less blue light contrasting the red and the $\acsi-\wfch$ color again shifts toward
the red. Finally, if a galaxy dominated by older stellar populations is dust attenuated then
the red stellar populations become more red, shifting the $\acsi-\wfch$ color, but the ICD
would increase as there is now some contrast between stellar populations.

In Figure~\ref{fig:icd_vs_color} we show the ICD as a function of $\acsi-\wfch$ color and
attenuation value ($\rm A_V$) as reported by {\tt FAST} (see Section~\ref{sec:Stellar
Masses}). There is little evidence for a correlations between either the ICD or $\rm A_V$ as
we find high $\xi(\acsi, \wfch)$ galaxies for a wide range of colors and $\rm A_V$. The ICD
is low for galaxies with both high and low $\acsi - \wfch$ colors, but galaxies with
moderate $\acsi - \wfch$ color (1 mag $< \acsi - \wfch <$ 2 mag) show the largest range in
the ICD. Therefore, it seems that galaxies with moderate $\acsi - \wfch$ colors are required
for high ICD, but color alone is not enough to predict high ICD. \cite{Papovich2005,
Bond2011} argue this lack of direct correlation supports the idea that the ICD depends on
spatial variations in the ages of the stellar populations and not dust attenuation.

\begin{figure} \includegraphics[width=0.49\textwidth]{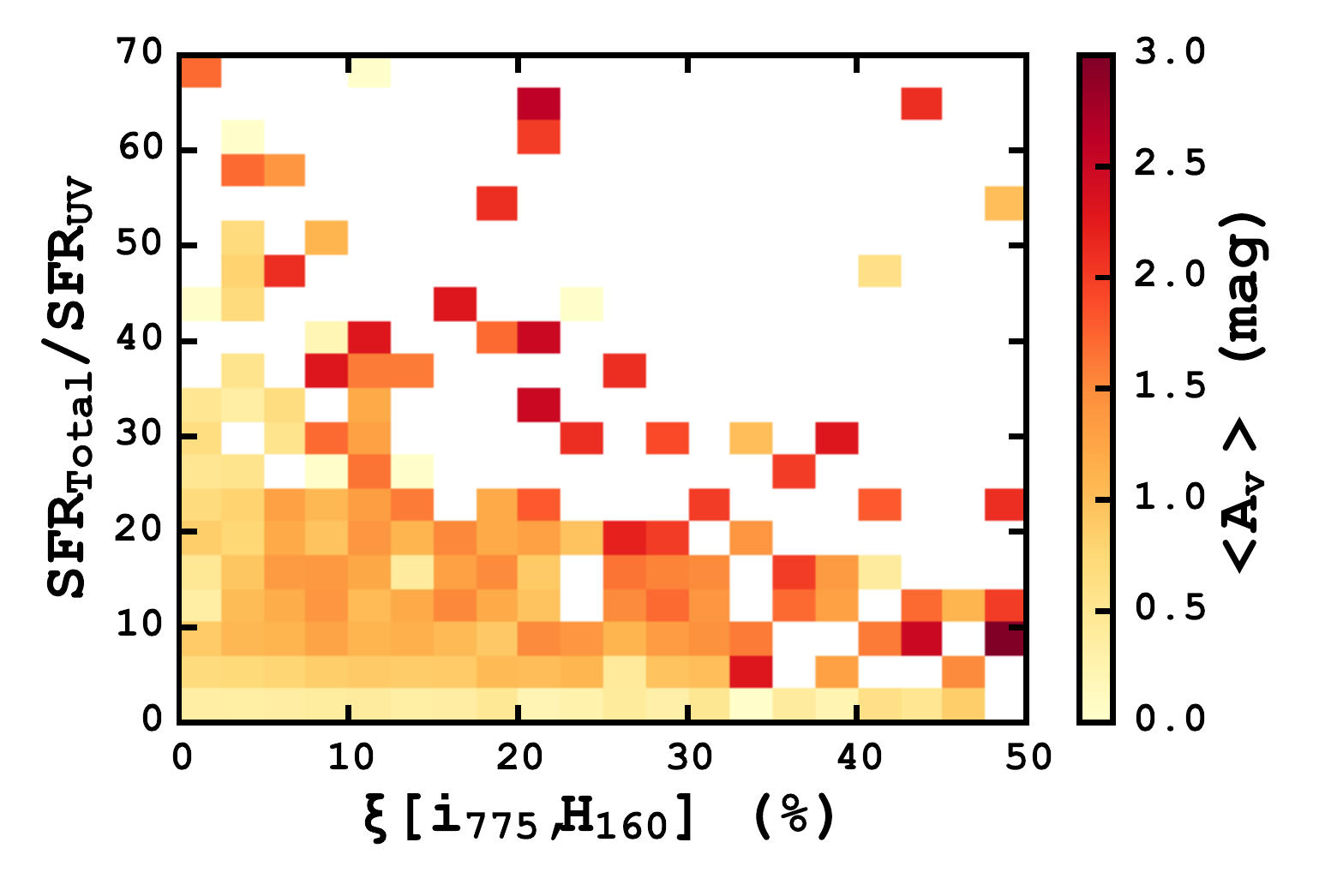} \caption{The ratio of the total SFR to the SFR derived from observed UV luminosity, $\rm SFR_{Total}/SFR_{UV}$, as a function of ICD. The colored boxes show the average attenuation value ($\rm A_V$) of the points within the region. Galaxies with IR dominated SFRs have smaller ICDs than those with a lower $\rm SFR_{Total}/SFR_{UV}$. At fixed $\rm SFR_{Total}/SFR_{UV}$, there is a large spread in derived attenuation. This suggests that a mix of stellar population ages drives the ICD and attenuation plays a minor role.} \label{fig:sfrRatio_icd} \end{figure}

As a second investigation, we compare the ratio of the total galaxy SFR ($\rm SFR_{total}$),
derived from the IR and UV emission (see Section~\ref{sec:data_sfr}), against the SFR
inferred from the UV measurement (uncorrected for dust attenuation). In this way, the ratio
of $\rm SFR_{total}/SFR_{UV}$ is a measure of the amount of light from star formation that
is attenuated by dust and reradiated in the IR. Figure~\ref{fig:sfrRatio_icd} shows there is
no trend in $\rm SFR_{total}/SFR_{UV}$ as a function of $\xi(\acsi, \wfch)$. Galaxies with
high $\xi(\acsi, \wfch)$ span the entire range of $\rm SFR_{total}/SFR_{UV}$.

Taken together, this is evidence that spatial variations of dust is not the main driver of
the ICD, and as such, the ICD requires spatial variations in stellar population ages in
galaxies (perhaps with contributions from spatial variations in dust attenuation). This
agrees with the conclusions from \cite{Papovich2003, Papovich2005} and \cite{Bond2011}. Both
authors acknowledge that extinction variations could play a role in providing variations in
the blue starlight, but conclude that the effect is smaller than those produced by
variations in segregated stellar populations of different mass-weighted ages. The ICD
appears to driven mostly by variations in the ages of stellar populations in galaxies.

\subsection{Star Forming Clumps}\label{sec:clumps} \cite{Papovich2005} suggest that in order
to have ICD $>10\%$, young star-forming disk in galaxies are required to have a surrounding
older spheroid; for higher ICDs ($>20-25\%$), they suggest compact star-forming HII regions
(discrete star-forming regions or clumps) are also required. This picture is corroborated in
a recent study by \cite{Wuyts2012} who found that star-forming galaxies at $z\sim2$, with
stellar masses $\mathrm{Log~M} / \Msol > 10$ have redder centers and bluer outer regions.

\begin{figure} \includegraphics[width=0.49\textwidth]{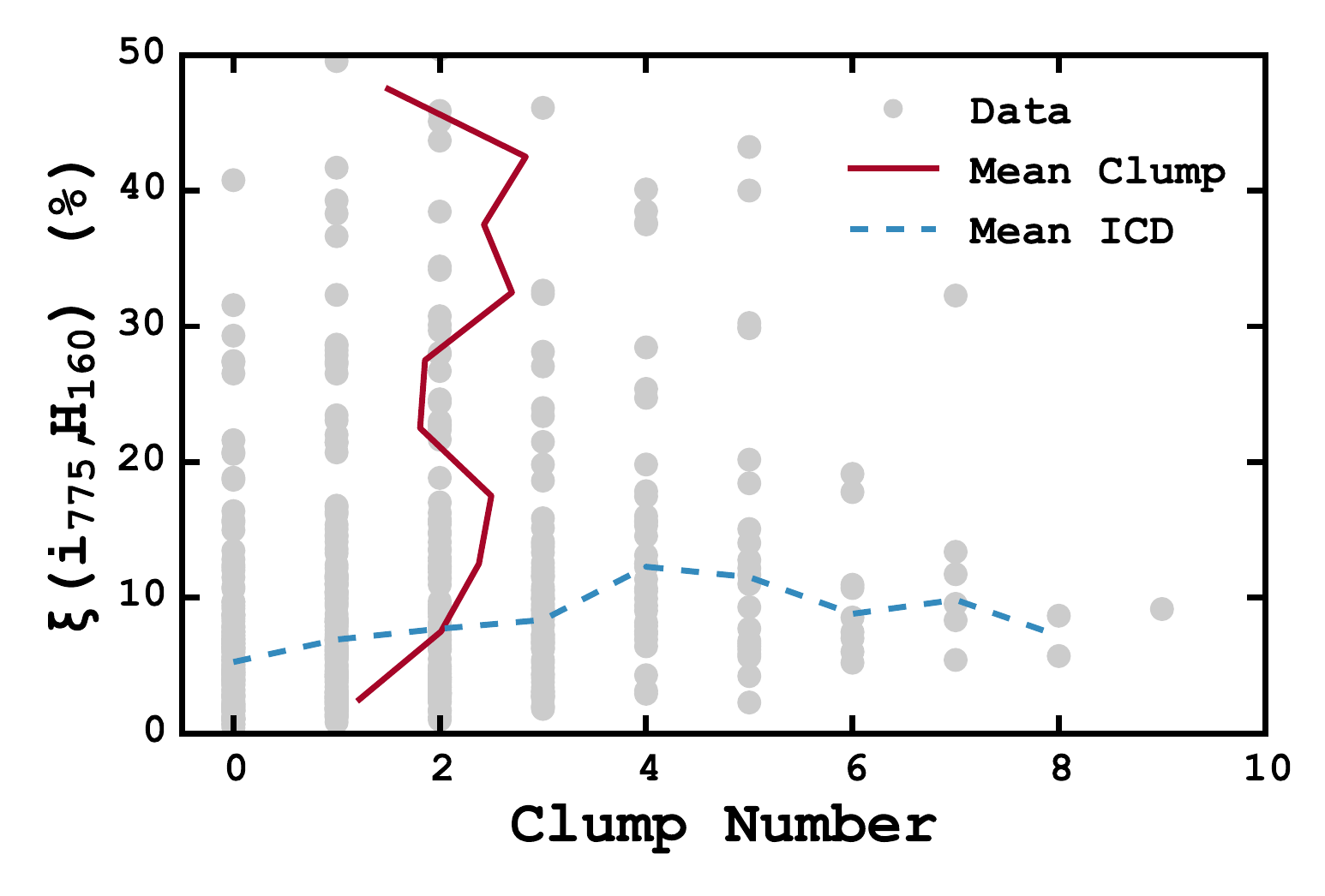} \caption{ICD as a function of number of clumps. The gray points are the individual galaxies. The red solid and blue dashed lines represent the average clump number in ICD bins and the average ICD value for different
numbers of clumps.} \label{fig:clump_icd} \end{figure}

Figure~\ref{fig:clump_icd} shows $\xi(\acsi, \wfch)$ as a function of the number of clumps.
using the catalog of \cite{Guo2014}, see Section~\ref{sec:data_clumps}. The average
$\xi(\acsi, \wfch)$ value steadily increases for increasing number of clumps but turns over
after a galaxy contains at least four clumps. The average number of clumps for a given
$\xi(\acsi, \wfch)$ value shows a similar increase with increasing $\xi(\acsi, \wfch)$ but
does not show a turnover. For galaxies with low ($<5\%$) ICD values, the mean number of
clumps per galaxy is 1.2; for galaxies with higher ICDs ($>5\%$), the mean number of clumps
is 2.1 Therefore, we find that if a galaxy has at least two clumps, on the average, it will
have a high $\xi(\acsi, \wfch)$ value. This trend continues, where galaxies with four or
more clumps show median ICD of greater than 10\%. Therefore, the ICD is sensitive to the number of star-forming clumps.

The clump catalog provides several clump detection levels (see
Section~\ref{sec:data_clumps}). When we increase in the detection threshold (UV luminosity
$>5\%$ instead of $1\%$) we find the mean number of clumps remains roughly constant ($1-2$)
regardless of the $\xi(\acsi, \wfch)$ value. Also, while the total number of clumps per
galaxy is reduced, we find the $\xi(\acsi, \wfch)$ of galaxies with fewer than two clumps
remains generally unchanged between the two UV luminosity thresholds.

Because the star-forming clumps are expected to form in gas-rich disks (\eg\
\citealt{Keres2005, Dekel2006, Dekel2009a, Dekel2014}), the ICD must trace galaxies which
have a redder but still gaseous stellar disk. These individual clumps are thought to be
relatively short lived (only $100 -200$ Myr; \citealt{Wuyts2012}), which would correspond
well with the lifetimes of hot, blue stars contributing to the UV light associated with high
ICD.

%%%%%%%%%%%%%%%%%%%%%%%% %%% BEGIN DISCUSSION %%% %%%%%%%%%%%%%%%%%%%%%%%%

\section{On Bulges, Clumps, and the ICD}\label{sec:On Morphology} For the galaxies in this
study, we find that as both the stellar mass and S\'{e}rsic index increase, the ICD
decreases (Figure~\ref{fig:icd_vs_mass_vs_sersic}). This is unsurprising as previous studies
suggest galaxies with high S\'{e}rsic indices have smooth light profiles with little color
variation across them (\eg\ \citealt{VanDokkum2004, Daddi2004, Szomoru2010}) and these
stellar populations produce little color dispersion as any variation is small relative to
the total galaxy flux \citep{Papovich2005}.

However, the galaxies with high ICD values have intermediate stellar mass and are
disk-dominated ($n<2$). As mentioned previously, \cite{Papovich2005} found that for a galaxy
to have high ICD a red bulge surrounded by a blue disk with HII regions (\ie\ clumps) is
required. Because high ICD is most prominent in disk-dominated morphologies and associated
with star-forming clumps, the ICD is less sensitive to the red bulge component and more
driven by the color variation between the underlying stellar disk and highly star-forming
clump. We see this in the previous comparison of the $\xi(\acsi, \wfch)$ computed with and
without the central region (see Section~\ref{sec:Color Gradients}), where we found that
galaxies with their nuclear region removed show an increased $\xi(\acsi, \wfch)$
(Figure~\ref{fig:icd_icdcored}).

Interpreted another way, the decrease of $\xi(\acsi, \wfch)$ with increasing S\'{e}rsic
index suggests that there is a continual suppression of the mechanism which causes high ICD
(perhaps clumps) with spheroid formation. This shows that \textit{if} galaxies are to form
quiescent bulges or spheroids along with a star-forming disk or clumps (as is typical of
classical Hubble Sequence galaxies) it is only happening at this moderate ($10 <
\mathrm{Log~M} / \Msol < 11$) stellar mass range and with relatively small bulge
contributions. Once a bulge contributes significantly to the galaxy's morphology, the amount
of ICD decreases dramatically, suggesting that the bulge plays a role in the homogenization
of the galaxy's stellar populations.

Two different effects could be associated with the formation of a bulge and subsequent
reduction in the observed ICD. One theory is the ``morphological quenching''
\citep{Martig2009} concept, by which the formation of the a bulge stabilizes the gaseous
disk against further fracturing and star formation. The second could be some feedback
process associated with the bulge which removes any remaining gas from the disk, preventing
new clump formation.

The first step in the quenching process is to form a strong bulge component. One scenario
for this process is for clumps to migrate through dynamical friction to coalesce in the
nuclei, and merge to form a bulge (\eg, \citealt{Elmegreen2008, Dekel2009a, Dekel2009b,
Ceverino2010}). The galaxies with high ICD values also show negative color gradients
(Figure~\ref{fig:icd_vs_color_grad}) suggesting the central stars are older (or perhaps more
attenuated by dust) than the outer star-forming regions. Other studies (\eg,
\citealt{Guo2012, Wuyts2012, Wuyts2013, Adamo2013}) find that central clumps are both older
(by $>\sim500$ Myrs) and more attenuated than the clumps in the outskirts suggesting that
migration plays a significant role.

Of course, not all galaxies in the moderate stellar mass range have high ICD values. Either
only some galaxies have periods of high ICD or all galaxies go through such a phase, but the
period of elevated ICD is very short lived. If all galaxies experience a period of elevated
ICD at some point during their evolution, then the characteristic time scale of high ICD is
the product of the time spanned by the look-back time across the redshift range (2.4 Grys)
and the fraction of sources with high ICD. In this work, 21\% of galaxies have $\xi(\acsi,
\wfch) > 10\%$. Therefore, if all galaxies go through a high ICD phase, then the timescale
of high ICD is approximately 375 Myrs. This timescale is similar to the predicted clump
survival timescale of $200-250$ Myrs \citep{Wuyts2012}.

Regardless of the mechanism at work, there is a distinct suppression in the ICD of galaxies
once a bulge is formed. Galaxies which contain highly star forming clumps match well the
description of these high ICD objects and clump migration could be a primary cause of ICD
suppression. It will take further investigation in the future to better associate galaxies
showing high ICD with star forming clumps.

\section{SUMMARY}\label{sec:summary} In this work, we investigate the dependence of the
rest-frame UV and optical ICD on other galaxy properties. We define the ICD as the ratio of
the square difference of the image flux -- intensity values about the mean galaxy color to
the square of the total image flux. It is a measure of the amount of variation in the
stellar diversity of a galaxy; that is separated red and blue stellar populations. Using
imaging from CANDELS of GOODS-South and the HUDF, we select galaxies between $1.5 <z< 3.5$,
a \wfch\ magnitude brighter than 25 mag and a requirement that the signal-to-noise be
sufficient enough to provide an accurate measurement of the ICD (see Section~\ref{sec:QC and
selection}). This selection yields 2601 galaxies.

We present our results from comparing the ICD to the stellar mass of the galaxies
(Section~\ref{sec:icdvmass}), the morphology based on S\'{e}risc index
(Section~\ref{sec:sersic}) and star formation rates (Section~\ref{sec:icdvssfr}). We then
investigate possible causes of the ICD including: mergers (Section~\ref{sec:mergers}),
physical size (Section~\ref{sec:size}), color gradients (Section~\ref{sec:Color Gradients}),
band-pass shifting effects (Section~\ref{sec:redshift}), attenuation
(Section~\ref{sec:attenuation}) and clumpy star forming regions (Section~\ref{sec:clumps}).
We attribute ICD to heterogeneous stellar populations (possibly clumps) with a possible
contribution from dust attenuation.

The major results of this work, with accompanying interpretations, are as follows:

\begin{itemize} \item Galaxies have the largest amount of stellar population diversity, as
measured by the ICD, at stellar masses $10 < \mathrm{Log~M}/\Msol < 11$. This stellar mass
range corresponds well to the range in which local galaxies experience a decrease in their
star formation efficiency, and galaxies at $z\sim2$ which are found to show star forming
clumps. Therefore, the stellar population diversity driving the ICD could be discrete,
highly star-forming regions contrasted against an older, redder, underlying population.

\item High ICD is preferentially found in galaxies with extended light profiles (S\'{e}risc
index $< 2$). This suggests that galaxies only have diverse stellar populations when there
is little or no bulge present.

\item There is a continual suppression in the mechanism which causes high ICD with
increasing S\'{e}risc index. This could be due formation of clumps, which initially increase
the ICD, and then migrate toward the center and coalescence into a quiescent bulge. The
presence of this bulge then stabilizes the disk again fragmentation and new clump formation,
which suppresses the ICD. \end{itemize}

The results suggests that some fraction of galaxies with high ICD are a result of the
combination of structures such as quiescent bulges and star-forming (perhaps clump) disks in
galaxies. If this corresponds to the formation of the spiral galaxies indicative of the
Hubble Sequence, then this phase occurs only in galaxies with stellar masses $10 <
\mathrm{Log~M}/\Msol < 11$ and when the bulge component remains relatively small ($n<2$).

\acknowledgments The authors also wish to thank the anonymous referee whose comments and
suggestions significantly improved both the quality and clarity of this work. This work is
based on observations taken by the CANDELS Multi-Cycle Treasury Program with the NASA/ESA
HST, which is operated by the Association of Universities for Research in Astronomy, Inc.,
under NASA contract NAS5-26555. We also make use of partial support from NSF AST-0808133. We
utilize the Rainbow Cosmological Surveys Database, which is operated by the Universidad
Complutense de Madrid (UCM), partnered with the University of California Observatories at
Santa Cruz (UCO/Lick,UCSC). Several open source resources are used to complete this study
Python \citep{van1991} along with Matplotlib \citep{Hunter2007} and IPython
\citep{Perez2007}.

%shoutout to @OverheardOnAph. Glad this is over.

%\bibliography{../master}

\begin{thebibliography}{}
\expandafter\ifx\csname natexlab\endcsname\relax\def\natexlab#1{#1}\fi

\bibitem[{Abraham {et~al.}(2007)Abraham, Nair, McCarthy, Glazebrook, Mentuch,
  Yan, Savaglio, Crampton, Murowinski, Juneau, {Le Borgne}, Carlberg,
  Jorgensen, Roth, Chen, \& Marzke}]{Abraham2007}
Abraham, R.~G., Nair, P., McCarthy, P.~J., {et~al.} 2007, The Astrophysical
  Journal, 669, 184

\bibitem[{Adamo {et~al.}(2013)Adamo, \"{O}stlin, Bastian, Zackrisson,
  Livermore, \& Guaita}]{Adamo2013}
Adamo, A., \"{O}stlin, G., Bastian, N., {et~al.} 2013, The Astrophysical
  Journal, 766, 105

\bibitem[{Arnouts {et~al.}(1999)Arnouts, Cristiani, Moscardini, Matarrese,
  Lucchin, Fontana, \& Giallongo}]{Arnouts1999}
Arnouts, S., Cristiani, S., Moscardini, L., {et~al.} 1999, Monthly Notices of
  the Royal Astronomical Society, 310, 540

\bibitem[{Baldry {et~al.}(2004)Baldry, Glazebrook, Brinkmann, Ivezi\'{c},
  Lupton, Nichol, \& Szalay}]{Baldry2004}
Baldry, I.~K., Glazebrook, K., Brinkmann, J., {et~al.} 2004, The Astrophysical
  Journal, 600, 681

\bibitem[{Beckwith {et~al.}(2006)Beckwith, Stiavelli, Koekemoer, Caldwell,
  Ferguson, Hook, Lucas, Bergeron, Corbin, Jogee, Panagia, Robberto, Royle,
  Somerville, \& Sosey}]{Beckwith2006}
Beckwith, S. V.~W., Stiavelli, M., Koekemoer, A.~M., {et~al.} 2006, The
  Astronomical Journal, 132, 1729

\bibitem[{Bell {et~al.}(2012)Bell, van~der Wel, Papovich, Kocevski, Lotz,
  McIntosh, Kartaltepe, Faber, Ferguson, Koekemoer, Grogin, Wuyts, Cheung,
  Conselice, Dekel, Dunlop, Giavalisco, Herrington, Koo, McGrath, de~Mello,
  Rix, Robaina, \& Williams}]{Bell2012}
Bell, E.~F., van~der Wel, A., Papovich, C., {et~al.} 2012, The Astrophysical
  Journal, 753, 167

\bibitem[{Bertin \& Arnouts(1996)}]{Bertin1996}
Bertin, E., \& Arnouts, S. 1996, Astronomy and Astrophysics Supplement Series,
  117, 393

\bibitem[{Blanton {et~al.}(2001)Blanton, Dalcanton, Eisenstein, Loveday,
  Strauss, SubbaRao, Weinberg, {Anderson, Jr.}, Annis, Bahcall, Bernardi,
  Brinkmann, Brunner, Burles, Carey, Castander, Connolly, Csabai, Doi,
  Finkbeiner, Friedman, Frieman, Fukugita, Gunn, Hennessy, Hindsley, Hogg,
  Ichikawa, Ivezi\'{c}, Kent, Knapp, Lamb, Leger, Long, Lupton, McKay, Meiksin,
  Merelli, Munn, Narayanan, Newcomb, Nichol, Okamura, Owen, Pier, Pope,
  Postman, Quinn, Rockosi, Schlegel, Schneider, Shimasaku, Siegmund, Smee,
  Snir, Stoughton, Stubbs, Szalay, Szokoly, Thakar, Tremonti, Tucker, Uomoto,
  {Vanden Berk}, Vogeley, Waddell, Yanny, Yasuda, \& York}]{Blanton2001a}
Blanton, M.~R., Dalcanton, J., Eisenstein, D., {et~al.} 2001, The Astronomical
  Journal, 121, 2358

\bibitem[{Blanton {et~al.}(2003)Blanton, Hogg, Bahcall, Baldry, Brinkmann,
  Csabai, Eisenstein, Fukugita, Gunn, Ivezi\'{c}, Lamb, Lupton, Loveday, Munn,
  Nichol, Okamura, Schlegel, Shimasaku, Strauss, Vogeley, \&
  Weinberg}]{Blanton2003a}
Blanton, M.~R., Hogg, D.~W., Bahcall, N.~A., {et~al.} 2003, The Astrophysical
  Journal, 594, 186

\bibitem[{Bolzonella {et~al.}(2000)Bolzonella, Miralles, \&
  Pell\'{o}}]{Bolzonella2000}
Bolzonella, M., Miralles, J.-M., \& Pell\'{o}, R. 2000, Astronomy and
  Astrophysics

\bibitem[{Bond {et~al.}(2011)Bond, Gawiser, \& Koekemoer}]{Bond2011}
Bond, N.~a., Gawiser, E., \& Koekemoer, A.~M. 2011, The Astrophysical Journal,
  729, 48

\bibitem[{Bouwens {et~al.}(2010)Bouwens, Illingworth, Oesch, Trenti, Stiavelli,
  Carollo, Franx, van Dokkum, Labb\'{e}, \& Magee}]{Bouwens2010}
Bouwens, R.~J., Illingworth, G.~D., Oesch, P.~A., {et~al.} 2010, The
  Astrophysical Journal, 708, L69

\bibitem[{Brammer {et~al.}(2008)Brammer, van Dokkum, \& Coppi}]{Brammer2008}
Brammer, G.~B., van Dokkum, P.~G., \& Coppi, P. 2008, The Astrophysical
  Journal, 686, 1503

\bibitem[{Bruzual \& Charlot(2003)}]{Bruzual2003}
Bruzual, G., \& Charlot, S. 2003, Monthly Notices of the Royal Astronomical
  Society, 344, 1000

\bibitem[{Calzetti {et~al.}(2000)Calzetti, Armus, Bohlin, Kinney, Koornneef, \&
  Storchi‚ÄêBergmann}]{Calzetti2000}
Calzetti, D., Armus, L., Bohlin, R.~C., {et~al.} 2000, The Astrophysical
  Journal, 533, 682

\bibitem[{Ceverino {et~al.}(2010)Ceverino, Dekel, \& Bournaud}]{Ceverino2010}
Ceverino, D., Dekel, A., \& Bournaud, F. 2010, Monthly Notices of the Royal
  Astronomical Society, 404, 2151

\bibitem[{Chabrier(2003)}]{Chabrier2003}
Chabrier, G. 2003, Publications of the Astronomical Society of the Pacific,
  115, 763

\bibitem[{Conselice {et~al.}(2004)Conselice, Grogin, Jogee, Lucas, Dahlen,
  de~Mello, Gardner, Mobasher, \& Ravindranath}]{Conselice2004}
Conselice, C.~J., Grogin, N.~A., Jogee, S., {et~al.} 2004, The Astrophysical
  Journal, 600, L139

\bibitem[{Daddi {et~al.}(2004)Daddi, Cimatti, Renzini, Fontana, Mignoli,
  Pozzetti, Tozzi, \& Zamorani}]{Daddi2004}
Daddi, E., Cimatti, A., Renzini, A., {et~al.} 2004, The Astrophysical Journal,
  617, 746

\bibitem[{Daddi {et~al.}(2007)Daddi, Dickinson, Morrison, Chary, Cimatti,
  Elbaz, Frayer, Renzini, Pope, Alexander, Bauer, Giavalisco, Huynh, Kurk, \&
  Mignoli}]{Daddi2007a}
Daddi, E., Dickinson, M., Morrison, G., {et~al.} 2007, The Astrophysical
  Journal, 670, 156

\bibitem[{Dahlen {et~al.}(2013)Dahlen, Mobasher, Faber, Ferguson, Barro,
  Finkelstein, Finlator, Fontana, Gruetzbauch, Johnson, Pforr, Salvato,
  Wiklind, Wuyts, Acquaviva, Dickinson, Guo, Huang, Huang, Newman, Bell,
  Conselice, Galametz, Gawiser, Giavalisco, Grogin, Hathi, Kocevski, Koekemoer,
  Koo, Lee, McGrath, Papovich, Peth, Ryan, Somerville, Weiner, \&
  Wilson}]{Dahlen2013}
Dahlen, T., Mobasher, B., Faber, S.~M., {et~al.} 2013, The Astrophysical
  Journal, 775, 93

\bibitem[{de~Vaucouleurs(1948)}]{DeVaucouleurs1948}
de~Vaucouleurs, G. 1948, Annales d'Astrophysique, 11

\bibitem[{Dekel \& Birnboim(2006)}]{Dekel2006}
Dekel, A., \& Birnboim, Y. 2006, Monthly Notices of the Royal Astronomical
  Society, 368, 2

\bibitem[{Dekel \& Burkert(2014)}]{Dekel2014}
Dekel, A., \& Burkert, A. 2014, Monthly Notices of the Royal Astronomical
  Society, 438, 1870

\bibitem[{Dekel {et~al.}(2009{\natexlab{a}})Dekel, Sari, \&
  Ceverino}]{Dekel2009b}
Dekel, A., Sari, R., \& Ceverino, D. 2009{\natexlab{a}}, The Astrophysical
  Journal, 703, 785

\bibitem[{Dekel {et~al.}(2009{\natexlab{b}})Dekel, Birnboim, Engel, Freundlich,
  Goerdt, Mumcuoglu, Neistein, Pichon, Teyssier, \& Zinger}]{Dekel2009a}
Dekel, A., Birnboim, Y., Engel, G., {et~al.} 2009{\natexlab{b}}, Nature, 457,
  451

\bibitem[{Dickinson(2000)}]{Dickinson2000}
Dickinson, M. 2000, Philosophical Transactions of the Royal Society A:
  Mathematical, Physical and Engineering Sciences, 358, 2001

\bibitem[{Dickinson {et~al.}(2003)Dickinson, Papovich, Ferguson, \&
  Budavari}]{Dickinson2003a}
Dickinson, M., Papovich, C., Ferguson, H.~C., \& Budavari, T. 2003, The
  Astrophysical Journal, 587, 25

\bibitem[{Dutton(2009)}]{Dutton2009}
Dutton, A.~A. 2009, Monthly Notices of the Royal Astronomical Society, 396, 121

\bibitem[{Ellis {et~al.}(2013)Ellis, McLure, Dunlop, Robertson, Ono, Schenker,
  Koekemoer, Bowler, Ouchi, Rogers, Curtis-Lake, Schneider, Charlot, Stark,
  Furlanetto, \& Cirasuolo}]{Ellis2013}
Ellis, R.~S., McLure, R.~J., Dunlop, J.~S., {et~al.} 2013, The Astrophysical
  Journal, 763, L7

\bibitem[{Elmegreen {et~al.}(2008)Elmegreen, Bournaud, \&
  Elmegreen}]{Elmegreen2008}
Elmegreen, B.~G., Bournaud, F., \& Elmegreen, D.~M. 2008, The Astrophysical
  Journal, 688, 67

\bibitem[{Elmegreen {et~al.}(2009{\natexlab{a}})Elmegreen, Elmegreen,
  Fernandez, \& Lemonias}]{Elmegreen2009}
Elmegreen, B.~G., Elmegreen, D.~M., Fernandez, M.~X., \& Lemonias, J.~J.
  2009{\natexlab{a}}, The Astrophysical Journal, 692, 12

\bibitem[{Elmegreen {et~al.}(2009{\natexlab{b}})Elmegreen, Elmegreen, Marcus,
  Shahinyan, Yau, \& Petersen}]{Elmegreen2009a}
Elmegreen, D.~M., Elmegreen, B.~G., Marcus, M.~T., {et~al.} 2009{\natexlab{b}},
  The Astrophysical Journal, 701, 306

\bibitem[{Elmegreen {et~al.}(2005)Elmegreen, Elmegreen, Rubin, \&
  Schaffer}]{Elmegreen2005}
Elmegreen, D.~M., Elmegreen, B.~G., Rubin, D.~S., \& Schaffer, M.~A. 2005, The
  Astrophysical Journal, 631, 85

\bibitem[{{F\"{o}rster Schreiber} {et~al.}(2011){F\"{o}rster Schreiber},
  Shapley, Genzel, Bouch\'{e}, Cresci, Davies, Erb, Genel, Lutz, Newman,
  Shapiro, Steidel, Sternberg, \& Tacconi}]{ForsterSchreiber2011}
{F\"{o}rster Schreiber}, N.~M., Shapley, A.~E., Genzel, R., {et~al.} 2011, The
  Astrophysical Journal, 739, 45

\bibitem[{Franx {et~al.}(2008)Franx, van Dokkum, Schreiber, Wuyts, Labb\'{e},
  \& Toft}]{Franx2008}
Franx, M., van Dokkum, P.~G., Schreiber, N. M.~F., {et~al.} 2008, The
  Astrophysical Journal, 688, 770

\bibitem[{Giavalisco {et~al.}(1996)Giavalisco, Steidel, \&
  Macchetto}]{Giavalisco1996}
Giavalisco, M., Steidel, C.~C., \& Macchetto, F.~D. 1996, The Astrophysical
  Journal, 470, 189

\bibitem[{Giavalisco {et~al.}(2004)Giavalisco, Ferguson, Koekemoer, Dickinson,
  Alexander, Bauer, Bergeron, Biagetti, Brandt, Casertano, Cesarsky,
  Chatzichristou, Conselice, Cristiani, {Da Costa}, Dahlen, de~Mello,
  Eisenhardt, Erben, Fall, Fassnacht, Fosbury, Fruchter, Gardner, Grogin, Hook,
  Hornschemeier, Idzi, Jogee, Kretchmer, Laidler, Lee, Livio, Lucas, Madau,
  Mobasher, Moustakas, Nonino, Padovani, Papovich, Park, Ravindranath, Renzini,
  Richardson, Riess, Rosati, Schirmer, Schreier, Somerville, Spinrad, Stern,
  Stiavelli, Strolger, Urry, Vandame, Williams, \& Wolf}]{Giavalisco2004}
Giavalisco, M., Ferguson, H.~C., Koekemoer, A.~M., {et~al.} 2004, The
  Astrophysical Journal, 600, L93

\bibitem[{Grogin {et~al.}(2011)Grogin, Kocevski, Faber, Ferguson, Koekemoer,
  Riess, Acquaviva, Alexander, Almaini, Ashby, Barden, Bell, Bournaud, Brown,
  Caputi, Casertano, Cassata, Castellano, Challis, Chary, Cheung, Cirasuolo,
  Conselice, Cooray, Croton, Daddi, Dahlen, Dav\'{e}, de~Mello, Dekel,
  Dickinson, Dolch, Donley, Dunlop, Dutton, Elbaz, Fazio, Filippenko,
  Finkelstein, Fontana, Gardner, Garnavich, Gawiser, Giavalisco, Grazian, Guo,
  Hathi, H\"{a}ussler, Hopkins, Huang, Huang, Jha, Kartaltepe, Kirshner, Koo,
  Lai, Lee, Li, Lotz, Lucas, Madau, McCarthy, McGrath, McIntosh, McLure,
  Mobasher, Moustakas, Mozena, Nandra, Newman, Niemi, Noeske, Papovich,
  Pentericci, Pope, Primack, Rajan, Ravindranath, Reddy, Renzini, Rix, Robaina,
  Rodney, Rosario, Rosati, Salimbeni, Scarlata, Siana, Simard, Smidt,
  Somerville, Spinrad, Straughn, Strolger, Telford, Teplitz, Trump, van~der
  Wel, Villforth, Wechsler, Weiner, Wiklind, Wild, Wilson, Wuyts, Yan, \&
  Yun}]{Grogin2011}
Grogin, N.~a., Kocevski, D.~D., Faber, S.~M., {et~al.} 2011, The Astrophysical
  Journal Supplement Series, 197, 35

\bibitem[{Gr\"{u}tzbauch {et~al.}(2011)Gr\"{u}tzbauch, Chuter, Conselice,
  Bauer, Bluck, Buitrago, \& Mortlock}]{Grutzbauch2011}
Gr\"{u}tzbauch, R., Chuter, R.~W., Conselice, C.~J., {et~al.} 2011, Monthly
  Notices of the Royal Astronomical Society, 412, 2361

\bibitem[{Guo {et~al.}(2012)Guo, Giavalisco, Ferguson, Cassata, \&
  Koekemoer}]{Guo2012}
Guo, Y., Giavalisco, M., Ferguson, H.~C., Cassata, P., \& Koekemoer, A.~M.
  2012, The Astrophysical Journal, 757, 120

\bibitem[{Guo {et~al.}(2013)Guo, Ferguson, Giavalisco, Barro, Willner, Ashby,
  Dahlen, Donley, Faber, Fontana, Galametz, Grazian, Huang, Kocevski,
  Koekemoer, Koo, McGrath, Peth, Salvato, Wuyts, Castellano, Cooray, Dickinson,
  Dunlop, Fazio, Gardner, Gawiser, Grogin, Hathi, Hsu, Lee, Lucas, Mobasher,
  Nandra, Newman, \& van~der Wel}]{Guo2013}
Guo, Y., Ferguson, H.~C., Giavalisco, M., {et~al.} 2013, The Astrophysical
  Journal Supplement Series, 207, 24

\bibitem[{Guo {et~al.}(2014)Guo, Ferguson, Bell, Koo, Conselice, Giavalisco,
  Kassin, Lu, Lucas, Mandelker, McIntosh, Primack, Ravindranath, Barro,
  Ceverino, Dekel, Faber, Fang, Koekemoer, Noeske, Rafelski, \&
  Straughn}]{Guo2014}
Guo, Y., Ferguson, H.~C., Bell, E.~F., {et~al.} 2014, eprint arXiv:1410.7398,
  22

\bibitem[{Hunter(2007)}]{Hunter2007}
Hunter, J.~D. 2007, Computing in Science \& Engineering, 9, 90

\bibitem[{Ilbert {et~al.}(2006)Ilbert, Arnouts, McCracken, Bolzonella, Bertin,
  {Le F\`{e}vre}, Mellier, Zamorani, Pell\`{o}, Iovino, Tresse, {Le Brun},
  Bottini, Garilli, Maccagni, Picat, Scaramella, Scodeggio, Vettolani,
  Zanichelli, Adami, Bardelli, Cappi, Charlot, Ciliegi, Contini, Cucciati,
  Foucaud, Franzetti, Gavignaud, Guzzo, Marano, Marinoni, Mazure, Meneux,
  Merighi, Paltani, Pollo, Pozzetti, Radovich, Zucca, Bondi, Bongiorno,
  Busarello, {De La Torre}, Gregorini, Lamareille, Mathez, Merluzzi, Ripepi,
  Rizzo, \& Vergani}]{Ilbert2006}
Ilbert, O., Arnouts, S., McCracken, H.~J., {et~al.} 2006, Astronomy and
  Astrophysics, 457, 841

\bibitem[{Kauffmann {et~al.}(2003)Kauffmann, Heckman, White, Charlot, Tremonti,
  Peng, Seibert, Brinkmann, Nichol, SubbaRao, \& York}]{Kauffmann2003b}
Kauffmann, G., Heckman, T.~M., White, S. D.~M., {et~al.} 2003, Monthly Notices
  of the Royal Astronomical Society, 341, 54

\bibitem[{Kennicutt(1998)}]{Kennicutt1998}
Kennicutt, R.~C. 1998, Annual Review of Astronomy and Astrophysics, 36, 189

\bibitem[{Keres {et~al.}(2005)Keres, Katz, Weinberg, \& Dave}]{Keres2005}
Keres, D., Katz, N., Weinberg, D.~H., \& Dave, R. 2005, Monthly Notices of the
  Royal Astronomical Society, 363, 2

\bibitem[{Koekemoer {et~al.}(2011)Koekemoer, Faber, Ferguson, Grogin, Kocevski,
  Koo, Lai, Lotz, Lucas, McGrath, Ogaz, Rajan, Riess, Rodney, Strolger,
  Casertano, Castellano, Dahlen, Dickinson, Dolch, Fontana, Giavalisco,
  Grazian, Guo, Hathi, Huang, van~der Wel, Yan, Acquaviva, Alexander, Almaini,
  Ashby, Barden, Bell, Bournaud, Brown, Caputi, Cassata, Challis, Chary,
  Cheung, Cirasuolo, Conselice, Cooray, Croton, Daddi, Dav\'{e}, de~Mello,
  de~Ravel, Dekel, Donley, Dunlop, Dutton, Elbaz, Fazio, Filippenko,
  Finkelstein, Frazer, Gardner, Garnavich, Gawiser, Gruetzbauch, Hartley,
  H\"{a}ussler, Herrington, Hopkins, Huang, Jha, Johnson, Kartaltepe,
  Khostovan, Kirshner, Lani, Lee, Li, Madau, McCarthy, McIntosh, McLure,
  McPartland, Mobasher, Moreira, Mortlock, Moustakas, Mozena, Nandra, Newman,
  Nielsen, Niemi, Noeske, Papovich, Pentericci, Pope, Primack, Ravindranath,
  Reddy, Renzini, Rix, Robaina, Rosario, Rosati, Salimbeni, Scarlata, Siana,
  Simard, Smidt, Snyder, Somerville, Spinrad, Straughn, Telford, Teplitz,
  Trump, Vargas, Villforth, Wagner, Wandro, Wechsler, Weiner, Wiklind, Wild,
  Wilson, Wuyts, \& Yun}]{Koekemoer2011}
Koekemoer, A.~M., Faber, S.~M., Ferguson, H.~C., {et~al.} 2011, The
  Astrophysical Journal Supplement Series, 197, 36

\bibitem[{Kriek {et~al.}(2009)Kriek, van Dokkum, Labb\'{e}, Franx, Illingworth,
  Marchesini, \& Quadri}]{Kriek2009}
Kriek, M., van Dokkum, P.~G., Labb\'{e}, I., {et~al.} 2009, The Astrophysical
  Journal, 700, 221

\bibitem[{Laidler {et~al.}(2006)Laidler, Grogin, Clubb, Ferguson, Papovich,
  Dickinson, Idzi, MacDonald, Ouchi, \& Mobasher}]{Laidler2006}
Laidler, V.~G., Grogin, N., Clubb, K., {et~al.} 2006, Astronomical Data
  Analysis Software and Systems XV ASP Conference Series, 351

\bibitem[{Law {et~al.}(2007)Law, Steidel, Erb, Pettini, Reddy, Shapley,
  Adelberger, \& Simenc}]{Law2007}
Law, D.~R., Steidel, C.~C., Erb, D.~K., {et~al.} 2007, The Astrophysical
  Journal, 656, 1

\bibitem[{Law {et~al.}(2012)Law, Steidel, Shapley, Nagy, Reddy, \&
  Erb}]{Law2012}
Law, D.~R., Steidel, C.~C., Shapley, A.~E., {et~al.} 2012, The Astrophysical
  Journal, 745, 85

\bibitem[{Lee {et~al.}(2013)Lee, Giavalisco, Williams, Guo, Lotz, {Van der
  Wel}, Ferguson, Faber, Koekemoer, Grogin, Kocevski, Conselice, Wuyts, Dekel,
  Kartaltepe, \& Bell}]{Lee2013}
Lee, B., Giavalisco, M., Williams, C.~C., {et~al.} 2013, The Astrophysical
  Journal, 774, 47

\bibitem[{Lotz {et~al.}(2004)Lotz, Primack, \& Madau}]{Lotz2004a}
Lotz, J.~M., Primack, J., \& Madau, P. 2004, The Astronomical Journal, 128, 163

\bibitem[{Lotz {et~al.}(2008)Lotz, Davis, Faber, Guhathakurta, Gwyn, Huang,
  Koo, {Le Floc‚Äôh}, Lin, Newman, Noeske, Papovich, Willmer, Coil, Conselice,
  Cooper, Hopkins, Metevier, Primack, Rieke, \& Weiner}]{Lotz2008}
Lotz, J.~M., Davis, M., Faber, S.~M., {et~al.} 2008, The Astrophysical Journal,
  672, 177

\bibitem[{Lowenthal {et~al.}(1997)Lowenthal, Koo, Guzman, Gallego, Phillips,
  Faber, Vogt, Illingworth, \& Gronwall}]{Lowenthal1997}
Lowenthal, J.~D., Koo, D.~C., Guzman, R., {et~al.} 1997, The Astrophysical
  Journal, 481, 673

\bibitem[{Magdis {et~al.}(2010)Magdis, Rigopoulou, Huang, \&
  Fazio}]{Magdis2010a}
Magdis, G.~E., Rigopoulou, D., Huang, J.-S., \& Fazio, G.~G. 2010, Monthly
  Notices of the Royal Astronomical Society, 401, 1521

\bibitem[{Martig {et~al.}(2009)Martig, Bournaud, Teyssier, \&
  Dekel}]{Martig2009}
Martig, M., Bournaud, F., Teyssier, R., \& Dekel, A. 2009, The Astrophysical
  Journal, 707, 250

\bibitem[{Oesch {et~al.}(2010)Oesch, Bouwens, Illingworth, Carollo, Franx,
  Labb\'{e}, Magee, Stiavelli, Trenti, \& van Dokkum}]{Oesch2010}
Oesch, P.~A., Bouwens, R.~J., Illingworth, G.~D., {et~al.} 2010, The
  Astrophysical Journal, 709, L16

\bibitem[{Oke(1974)}]{Oke1974}
Oke, J.~B. 1974, The Astrophysical Journal Supplement Series, 27, 21

\bibitem[{Papovich {et~al.}(2001)Papovich, Dickinson, \&
  Ferguson}]{Papovich2001}
Papovich, C., Dickinson, M., \& Ferguson, H.~C. 2001, The Astrophysical
  Journal, 559, 620

\bibitem[{Papovich {et~al.}(2005)Papovich, Dickinson, Giavalisco, Conselice, \&
  Ferguson}]{Papovich2005}
Papovich, C., Dickinson, M., Giavalisco, M., Conselice, C.~J., \& Ferguson,
  H.~C. 2005, The Astrophysical Journal, 631, 101

\bibitem[{Papovich {et~al.}(2003)Papovich, Giavalisco, Dickinson, Conselice, \&
  Ferguson}]{Papovich2003}
Papovich, C., Giavalisco, M., Dickinson, M., Conselice, C.~J., \& Ferguson,
  H.~C. 2003, The Astrophysical Journal, 598, 827

\bibitem[{Papovich {et~al.}(2014)Papovich, Labb\'{e}, Quadri, Tilvi, Behroozi,
  Bell, Glazebrook, Spitler, Straatman, Tran, Cowley, Dav\'{e}, Dekel,
  Dickinson, Ferguson, Finkelstein, Gawiser, Inami, Faber, Kacprzak,
  Kawinwanchakij, Kocevski, Koekemoer, Koo, Kurczynski, Lotz, Lu, Lucas,
  McIntosh, Mehrtens, Mobasher, Monson, Morrison, Nanayakkara, Perrson, Salmon,
  Simons, Tomczak, van¬†Dokkum, Weiner, \& Willner}]{Papovich2014a}
Papovich, ¬., Labb\'{e}, ¬., Quadri, ¬., {et~al.} 2014, eprint arXiv:1412.3806

\bibitem[{Peng {et~al.}(2002)Peng, Ho, Impey, \& Rix}]{Peng2002}
Peng, C.~Y., Ho, L.~C., Impey, C.~D., \& Rix, H.-W. 2002, The Astronomical
  Journal, 124, 266

\bibitem[{Perez \& Granger(2007)}]{Perez2007}
Perez, F., \& Granger, B.~E. 2007, Computing in Science \& Engineering, 9, 21

\bibitem[{Salmon {et~al.}(2014)Salmon, Papovich, Finkelstein, Tilvi, Finlator,
  Behroozi, Dahlen, Dav\'{e}, Dekel, Dickinson, Ferguson, Giavalisco, Long, Lu,
  Reddy, Somerville, \& Wechsler}]{Salmon2014}
Salmon, B., Papovich, C., Finkelstein, S.~L., {et~al.} 2014, eprint
  arXiv:1407.6012

\bibitem[{Salpeter(1955)}]{Salpeter1955}
Salpeter, E.~E. 1955, The Astrophysical Journal, 121, 161

\bibitem[{S\'{e}rsic(1963)}]{Sersic1963}
S\'{e}rsic, J.~L. 1963, Boletin de la Asociacion Argentina de Astronomia, 6

\bibitem[{Speagle {et~al.}(2014)Speagle, Steinhardt, Capak, \&
  Silverman}]{Speagle2014a}
Speagle, J.~S., Steinhardt, C.~L., Capak, P.~L., \& Silverman, J.~D. 2014, The
  Astrophysical Journal Supplement Series, 214, 15

\bibitem[{Szomoru {et~al.}(2011)Szomoru, Franx, Bouwens, van Dokkum, Labb\'{e},
  Illingworth, \& Trenti}]{Szomoru2011}
Szomoru, D., Franx, M., Bouwens, R.~J., {et~al.} 2011, The Astrophysical
  Journal, 735, L22

\bibitem[{Szomoru {et~al.}(2010)Szomoru, Franx, van Dokkum, Trenti,
  Illingworth, Labb\'{e}, Bouwens, Oesch, \& Carollo}]{Szomoru2010}
Szomoru, D., Franx, M., van Dokkum, P.~G., {et~al.} 2010, The Astrophysical
  Journal, 714, L244

\bibitem[{van~der Wel {et~al.}(2012)van~der Wel, Bell, H\"{a}ussler, McGrath,
  Chang, Guo, McIntosh, Rix, Barden, Cheung, Faber, Ferguson, Galametz, Grogin,
  Hartley, Kartaltepe, Kocevski, Koekemoer, Lotz, Mozena, Peth, \&
  Peng}]{VanderWel2012}
van~der Wel, a., Bell, E.~F., H\"{a}ussler, B., {et~al.} 2012, The
  Astrophysical Journal Supplement Series, 203, 24

\bibitem[{van Dokkum {et~al.}(2004)van Dokkum, Franx, {Forster Schreiber},
  Illingworth, Daddi, Knudsen, Labbe, Moorwood, Rix, Rottgering, Rudnick,
  Trujillo, van~der Werf, van~der Wel, van Starkenburg, \&
  Wuyts}]{VanDokkum2004}
van Dokkum, P.~G., Franx, M., {Forster Schreiber}, N.~M., {et~al.} 2004, The
  Astrophysical Journal, 611, 703

\bibitem[{van Rossum \& de~Boer(1991)}]{van1991}
van Rossum, G., \& de~Boer, J. 1991, CWI Quarterly, 4, 283

\bibitem[{Wang {et~al.}(2012)Wang, Huang, Faber, Fang, Wuyts, Fazio, Yan,
  Dekel, Guo, Ferguson, Grogin, Lotz, Weiner, McGrath, Kocevski, Hathi, Lucas,
  Koekemoer, Kong, \& Gu}]{Wang2012}
Wang, T., Huang, J.-S., Faber, S.~M., {et~al.} 2012, The Astrophysical Journal,
  752, 134

\bibitem[{Williams {et~al.}(2010)Williams, Quadri, Franx, van Dokkum, Toft,
  Kriek, \& Labb\'{e}}]{Williams2010}
Williams, R.~J., Quadri, R.~F., Franx, M., {et~al.} 2010, The Astrophysical
  Journal, 713, 738

\bibitem[{Windhorst {et~al.}(2011)Windhorst, Cohen, Hathi, McCarthy, Ryan, Yan,
  Baldry, Driver, Frogel, Hill, Kelvin, Koekemoer, Mechtley, O'Connell,
  Robotham, Rutkowski, Seibert, Straughn, Tuffs, Balick, Bond, Bushouse,
  Calzetti, Crockett, Disney, Dopita, Hall, Holtzman, Kaviraj, Kimble,
  MacKenty, Mutchler, Paresce, Saha, Silk, Trauger, Walker, Whitmore, \&
  Young}]{Windhorst2011}
Windhorst, R.~A., Cohen, S.~H., Hathi, N.~P., {et~al.} 2011, The Astrophysical
  Journal Supplement Series, 193, 27

\bibitem[{Wuyts {et~al.}(2011{\natexlab{a}})Wuyts, {F\"{o}rster Schreiber},
  van~der Wel, Magnelli, Guo, Genzel, Lutz, Aussel, Barro, Berta, Cava,
  Graci\'{a}-Carpio, Hathi, Huang, Kocevski, Koekemoer, Lee, {Le Floc'h},
  McGrath, Nordon, Popesso, Pozzi, Riguccini, Rodighiero, Saintonge, \&
  Tacconi}]{Wuyts2011}
Wuyts, S., {F\"{o}rster Schreiber}, N.~M., van~der Wel, A., {et~al.}
  2011{\natexlab{a}}, The Astrophysical Journal, 742, 96

\bibitem[{Wuyts {et~al.}(2011{\natexlab{b}})Wuyts, {F\"{o}rster Schreiber},
  Lutz, Nordon, Berta, Altieri, Andreani, Aussel, Bongiovanni, Cepa, Cimatti,
  Daddi, Elbaz, Genzel, Koekemoer, Magnelli, Maiolino, McGrath, Garc\'{\i}a,
  Poglitsch, Popesso, Pozzi, Sanchez-Portal, Sturm, Tacconi, \&
  Valtchanov}]{Wuyts2011a}
Wuyts, S., {F\"{o}rster Schreiber}, N.~M., Lutz, D., {et~al.}
  2011{\natexlab{b}}, The Astrophysical Journal, 738, 106

\bibitem[{Wuyts {et~al.}(2012)Wuyts, {F\"{o}rster Schreiber}, Genzel, Guo,
  Barro, Bell, Dekel, Faber, Ferguson, Giavalisco, Grogin, Hathi, Huang,
  Kocevski, Koekemoer, Koo, Lotz, Lutz, McGrath, Newman, Rosario, Saintonge,
  Tacconi, Weiner, \& van~der Wel}]{Wuyts2012}
Wuyts, S., {F\"{o}rster Schreiber}, N.~M., Genzel, R., {et~al.} 2012, The
  Astrophysical Journal, 753, 114

\bibitem[{Wuyts {et~al.}(2013)Wuyts, {F\"{o}rster Schreiber}, Nelson, van
  Dokkum, Brammer, Chang, Faber, Ferguson, Franx, Fumagalli, Genzel, Grogin,
  Kocevski, Koekemoer, Lundgren, Lutz, McGrath, Momcheva, Rosario, Skelton,
  Tacconi, van~der Wel, \& Whitaker}]{Wuyts2013}
Wuyts, S., {F\"{o}rster Schreiber}, N.~M., Nelson, E.~J., {et~al.} 2013, The
  Astrophysical Journal, 779, 135

\bibitem[{Zibetti {et~al.}(2012)Zibetti, Gallazzi, Charlot, Pierini, \&
  Pasquali}]{Zibetti2012}
Zibetti, S., Gallazzi, A., Charlot, S., Pierini, D., \& Pasquali, A. 2012,
  Monthly Notices of the Royal Astronomical Society, 428, 1479

\end{thebibliography}

%\clearpage
%\begin{longtable*}{cccccccc}
%	\caption{Derived Quantities}\\
%	\hline
%ID & S/N \acsi\ & S/N \wfcj\ & S/N \acsv\ & $\xi(\acsi, \wfch)$ & $\xi(\wfcj, \wfch)$ & $\xi(\acsv, \wfcj)$ & Log Mass \\
%& & & & (\%) & (\%) & (\%) & (\Msol) \\
%	\hline \hline
%	\endfirsthead
%	\multicolumn{8}{c}%
%	{\tablename\ \thetable\ -- \textit{Continued from previous page}} \\
%	\hline
%ID & S/N \acsi\ & S/N \wfcj\ & S/N \acsv\ & $\xi(\acsi, \wfch)$ & $\xi(\wfcj, \wfch)$ & $\xi(\acsv, \wfcj)$ & Log %Mass \\
%& & & & (\%) & (\%) & (\%) & (\Msol) \\
%	\hline \hline
%	\endhead
%	\hline
%	\endfoot
%	\hline
%	\endlastfoot
%%%% THE TABLE GOES HERE %%%%
%\end{longtable*}
%\clearpage

\end{document}